\documentclass[bibyear]{aa}  

\usepackage{graphicx}
\usepackage{natbib}
\setcitestyle{aysep={}}
\usepackage{txfonts}
\usepackage{color}
\usepackage{tikz}
\def\ckmk{\tikz\fill[scale=0.4](0,.35) -- (.25,0) -- (1,.7) -- (.25,.15) -- cycle;}

\usepackage[colorlinks=true,citecolor=blue,linkcolor=blue]{hyperref}
\usepackage{lscape}

\begin{document}

   \title{The Belgian repository of fundamental atomic data and stellar spectra (BRASS). I. Cross-matching atomic databases of astrophysical interest}

   \author{M. Laverick\inst{1},
          A. Lobel\inst{2},
          T. Merle\inst{3},
          P. Royer\inst{1},
          C. Martayan\inst{4},
          M. David\inst{5},
          H. Hensberge\inst{2},
          and E. Thienpont\inst{6}
          }

   \institute{Instituut voor Sterrenkunde, KU~Leuven, Celestijnenlaan 200D, box 2401, 3001 Leuven, Belgium \\
                \email{mike.laverick@kuleuven.be}
         \and
             Royal Observatory of Belgium, Ringlaan 3, B-1180 Brussels, Belgium
         \and
              Institut d'Astronomie et d'Astrophysique, Universit\'{e} Libre de Bruxelles, Av. Franklin Roosevelt 50, CP 226, 1050 Brussels, Belgium
         \and
             European Organisation for Astronomical Research in the Southern Hemisphere, Alonso de C\'{o}rdova 3107, Vitacura, 19001 Casilla, Santiago de Chile, Chile
         \and
             University of Antwerp, Campus Middelheim, Middelheimlaan 1, 2020 Antwerp, Belgium,
         \and Vereniging voor Sterrenkunde, Kapellebaan 56, 2811 Leest, Belgium
             }

   \date{Received September 12, 2017; accepted December 15, 2017}

 
  \abstract
   {Fundamental atomic parameters, such as oscillator strengths, play a key role in modelling and understanding the chemical composition of stars in the universe. Despite the significant work underway to produce these parameters for many astrophysically important ions, uncertainties in these parameters remain large and can propagate throughout the entire field of astronomy.}
   {The Belgian repository of fundamental atomic data and stellar spectra (BRASS) aims to provide the largest systematic and homogeneous quality assessment of atomic data to date in terms of wavelength, atomic and stellar parameter coverage. To prepare for it, we first compiled multiple literature occurrences of many individual atomic transitions, from several atomic databases of astrophysical interest, and assessed their agreement. In a second step synthetic spectra will be compared against extremely high-quality observed spectra, for a large number of BAFGK spectral type stars, in order to critically evaluate the atomic data of a large number of important stellar lines.}
   {Several atomic repositories were searched and their data retrieved and formatted in a consistent manner. Data entries from all repositories were cross-matched against our initial BRASS atomic line list to find multiple occurrences of the same transition. Where possible we used a new non-parametric cross-match depending only on electronic configurations and total angular momentum values. We also checked for duplicate entries of the same physical transition, within each retrieved repository, using the non-parametric cross-match. }
   {We report on the number of cross-matched transitions for each repository and compare their fundamental atomic parameters. We find differences in $\log(gf)$ values of up to 2 dex or more. We also find and report that $\sim$2\% of our line list and Vienna Atomic Line Database retrievals are composed of duplicate transitions. Finally we provide a number of examples of atomic spectral lines with different retrieved literature $\log(gf)$ values, and discuss the impact of these uncertain $\log(gf)$ values on quantitative spectroscopy. All cross-matched atomic data and duplicate transition pairs are available to download at \url{brass.sdf.org}.   }
   {}
   \authorrunning{Laverick et al}
   \titlerunning{BRASS I. Cross-matching atomic databases of astrophysical interest}

   \maketitle

\section{Introduction}

Accurate atomic data are vital for stellar spectroscopy and understanding the chemical composition of the universe, and as such the importance of atomic data cannot be overstated. Errors in atomic data can systematically propagate throughout the entire field of astronomy and thus are frequently the subject of discussion \citep{basupergiants}, \citep{oscaccuracy}, \citep{lobelworkshop}, \citep{stellarlab}. Changes in both atomic data and spectral synthesis methods have led to significant revisions of solar abundances of elements that have had far reaching impacts on astronomy \citep{solar1998}, \citep{solar2008}, \citep{solar2009}, \citep{solar2011}, \citep{solar2015a}, \citep{solar2015b}, \citep{solar2015c}. Uncertainty in atomic data has led to several recent efforts, and many more ongoing, to update the atomic data of many ions with much-needed experimental data required for quantitative spectroscopy \citep{ti2update}, \citep{co1improve}, \citep{sc1update}, \citep{fe1update}, \citep{belmonte}. In addition, many more theoretical transitions are available due to increasingly more complex atomic calculations \citep{fe2calc}, \citep{ti2calc}, \citep{hf2calc}, \citep{mn2calc}, \citep{co2calc}. Online repositories such as the Vienna atomic line database, \citep{vald3}, the national institute of standards and technology atomic spectra database, \citep{nist}, and the providers within the  virtual atomic and molecular data centre, \citep{vamdc}, have made the retrieval of atomic line lists from the literature a much simpler task. Despite these recent efforts quality information for some transitions can remain poor or even absent from the literature, a problem only exacerbated by the ever increasing quantity of data. Previous efforts to constrain errors have typically employed a few benchmark stellar spectra of similar spectral type, with limited wavelength and ionic coverage \citep{valdupdate}, \citep{fe2bench}. 

The Belgian repository of fundamental atomic data and stellar spectra, BRASS \citep{alexbrass}, aims to provide the largest systematic and homogeneous quality assessment of atomic data to date in terms of wavelength, atomic species, and stellar parameter coverage. We shall critically evaluate the atomic data, such as wavelengths and oscillator strengths, of thousands of transitions found in the literature and across several major online atomic repositories. We shall compare synthetic spectrum calculations against extremely high-quality observed benchmark spectra for 20-30 objects spanning BAFGK spectral types. BRASS shall offer all quality assessed data, theoretical spectra, and observed spectra in a new interactive online database at \url{brass.sdf.org}. In addition to the 20-30 extremely high quality benchmark spectra, used for the atomic quality assessment, BRASS will provide over $\sim$100 high quality observed spectra of BAFGK type objects of astrophysical interest, including hot radial velocity standard stars (Lobel et al., in prep.). 

The compilation of accurate atomic line lists for large scale homogeneous quality assessments and stellar surveys, such as BRASS, the European space agency Gaia survey \citep{gaiamission}, and abundances of the APOGEE/Kepler sample \citep{apogeekepler}, is an arduous task and requires robust merging of data from multiple sources \citep{gaialines}. Given the far-reaching impact of large surveys any errors in line lists or uncertainties can easily propagate throughout the field and so it is of the utmost importance that line lists are built as accurately as possible. In this paper we shall describe the work of the BRASS project with regards to retrieving atomic data, cross-matching them, and exploring the agreement in atomic transition parameters amongst the literature. The cross-matches of previous work have typically relied on the similarity of atomic parameters, such as wavelengths and energy levels, to match similar transitions against each other \citep{vamandvald}. With this in mind we developed cross-matching tools, that use upper and lower electronic configurations to accurately match multiple occurrences of the same physical transition.

In Section 2 we discuss our retrieval of available atomic data from databases of astrophysical interest. In Section 3 we discuss our efforts to cross-match retrieved repositories and the details of both our parametric and non-parametric cross-match methods including inhomogeneities between the repositories. Section 4 explores the differences in atomic data between our successfully cross-matched transitions and the impact of such discrepancies. Section 5 describes our investigations into erroneous duplicate transitions present in line lists and the impact of these duplicates on synthetic spectra. Finally Section 6 provides examples of a number of stellar lines for which we found multiple literature occurrences of $\log(gf)$ values with significant variance. The examples clearly justify the need for a systematic  assessment of input atomic data to ensure that systematic errors in quantitative spectroscopy work are constrained.

\section{Retrieval of data from online repositories}
\begin{table*}
\centering
\caption{Number of retrieved lines, retrieval origin, retrieval date and available atomic data per repository. Wavelengths are in \aa ngstr\"{o}ms and energy in electron volts, eV. $i$ and $k$ indexes lower and upper states, respectively.}
\begin{tabular}{c c c c c c c c c c c c c}
            \hline
            \hline
            \noalign{\smallskip}
            Repository & Origin & No.~lines & Date &  Ion & $\lambda$ & $A_{ki}$ & $f_{ik}$ & $\log(gf)$& $E$ & $J$\\
            \noalign{\smallskip}
            \hline
            \noalign{\smallskip}
            BRASS		&-& 82337  & 2012~$^b$& \ckmk  &\ckmk      & 	    &		 &\ckmk     &\ckmk     &\ckmk \\
            SpectroWeb  &-& 62181  & 2008~$^b$& \ckmk  &\ckmk      & 	    &	     &\ckmk     &~~\ckmk$_f$ &	  \\
            VALD3 		&VALD& 158861 & 2016-05-26& \ckmk  &\ckmk      & 	    &	     &\ckmk     &\ckmk     &\ckmk \\
            NIST      	&NIST& 36123  & 2016-03-14& \ckmk  &~~\ckmk$_c$  &\ckmk &\ckmk  &\ckmk     &\ckmk     &\ckmk &\\
            Spectr-W$^3$   &VAMDC& 5515   & 2016-03-14  & \ckmk  &\ckmk      &\ckmk &\ckmk    &		    &\ckmk     &\ckmk \\
            TIPbase$^a$     &NORAD&33108   &2017-02-28 &\ckmk   & ~~\ckmk   &\ckmk  &\ckmk &	 &~~\ckmk$_g$& &      \\
            TOPbase$^a$     &VAMDC& 33462  & 2016-05-24& \ckmk  &~~\ckmk & 	    &		 &~~\ckmk$_e$ &~~\ckmk$_g$ &      \\
            CHIANTI    &VAMDC& 3587   & 2016-03-18& \ckmk  &~~\ckmk$_d$  &\ckmk & &~~\ckmk$_e$ &~~\ckmk$_f$ &\ckmk \\
            \noalign{\smallskip}
            \hline
\end{tabular}
\tablefoot{$^a$Data retrieved without wavelength restrictions and transitions are calculated without fine structure. $^b$Compiled from multiple sources over several months. $^c$Observed and Ritz wavelengths.  $^d$Vacuum wavelengths. $^e$Provided as $gf$ not $\log(gf)$. $^f$Energy in cm$^{-1}$. $^g$Energy in Ry.}
\label{numberoflines}
\end{table*}

In order to perform our atomic data quality assessment we have compiled an atomic line list (henceforth referred to as the BRASS atomic line list) which shall be used for all our spectral synthesis work. The BRASS atomic line list was compiled in 2012 using NIST v4.0 and the Kurucz website \citep{klines}, for the wavelength range 4200~-~6800~\AA~using ions of up to 5+. The BRASS atomic line list has been extensively tested in synthetic spectrum calculations against the solar Fourier transform spectrograph flux spectrum, described by \citet{solarfts}, and Mercator-HERMES \citep{hermes} spectra of various BAFGK stars, in combination with a variety of predicted di-atomic molecular line lists, discussed in \citet{alexbrass}. The BRASS atomic line list has been revised a number of times, removing unobserved spectral lines and spurious background features, as well as the removal of duplicated transitions discussed in Section 6. We also have access to the SpectroWeb \citep{spectroweb} atomic line lists which were compiled using VALD2 \citep{vald2} and NIST (v2.0 through v4.0), and extensively tested in a similar manner as the BRASS atomic line list. The SpectroWeb line list is, in essence, the predecessor of the BRASS atomic line list and comparisons between the two, and with newer versions of VALD and NIST, will allow us to track changes to the literature over time.
We gathered atomic data, in the wavelength range 4200-6800~\AA~and for ions up to 5+, from the following repositories \citep{nist}, \citep{vald3}, \citep{spectrw3}, \citep{tipbase}, \citep{topbase}, \citep{chianti}:

\begin{itemize}
\item NIST atomic spectra database - NIST ASD
\item Vienna atomic line database 3 - VALD3
\item Spectr-W$^3$ database
\item The iron project database - TIPbase
\item The opacity project database - TOPbase
\item CHIANTI atomic database
\end{itemize}

Table~\ref{numberoflines} shows the number of retrieved lines, retrieval origin, retrieval date and available atomic data for each of the databases and line lists. The majority of repositories were retrieved via the VAMDC. We are grateful for the current efforts of the VAMDC team in homogenising the repositories as this has helped to expedite up our comparisons and cross-match work. However it is worth noting, as shown in Table~\ref{numberoflines}, that the units of the different repositories are not homogeneous and care must be taken when working with different repositories at the same time. For the purposes of BRASS all retrieved data are homogenised prior to cross-matching.

\section{Cross-matching methods}

In order to cross-match two different occurrences of the same physical transition one would ideally search for transitions belonging to the same ion, with the same upper and lower electronic state configurations, same upper and lower total electron angular momentum \textit{J}-values, and the same upper and lower total atomic angular momentum \textit{F}-values. This should uniquely identify a given physical transition, where \textit{J}-values are used to distinguish fine structure and \textit{F}-values are used to distinguish hyperfine structure. We refer to this method of cross-matching as a non-parametric cross-match. 
A much simpler cross-match method is to search for transitions belonging to the same ion, with wavelengths, upper energies, and lower energies that all lie within a given threshold of each other.  We refer to this cross-match method as a parametric cross-match. This method can be further improved by requiring identical upper and lower parities between the transitions,  upper and lower \textit{J}-values and, if available, identical upper and lower \textit{F}-values. Fine structure information is typically available, but hyperfine structure information is limited and yet to be included in repositories. Such a method relies upon the assumption that the wavelengths and energies of two different occurrences of the same physical transition are similar, which is not always the case.

VALD3 and NIST both report lower and upper electronic configurations and terms in a similar and thorough manner meaning little to no manipulation, excluding the work to homogenise the parsing of configuration nomenclature, and thus we use the non-parametric cross-match method for these repositories. Spectr-W$^3$, TIPbase, TOPbase and CHIANTI all report on transition configurations, however they all require significant electronic state homogenisation before cross-matching. In this paper we opt to use the parametric cross-match for the Spectr-W$^3$ and CHIANTI repositories, and we discuss our efforts to cross-match TIPbase and TOPbase, which do not contain fine structure information.

As our BRASS atomic list configurations are compiled from both VALD and NIST, we are able to use the non-parametric cross-match for our work. While the SpectroWeb line list is also compiled from VALD and NIST, neither the electronic configurations nor the \textit{J}-values were retained, and so we are limited to the parametric cross-match in order to compare transitions. For the purposes of comparison we shall cross-match each of the retrieved repositories against our BRASS list. Each transition in the BRASS list is given a unique designation which will be used to keep track of all matched transition occurrences across the repositories. All cross-matched transitions including the BRASS atomic line list can be queried online, for a given wavelength range and elements, at \url{brass.sdf.org}.

\subsection{Non-parametric cross-match}

Through manual comparison of the differences between VALD and NIST electronic configuration nomenclature we found three major differences in the parsing of configurations that are inconsistently reported. These are listed below and examples of these differences are shown in Table~\ref{configdiff}:

\begin{enumerate}[(a)]
\item Closed sub-shells not always present
\item Lower case notation within the term and term seniority
\item Reporting the term within the configuration
\end{enumerate}

\begin{table*}
\centering
\caption{Examples of cross-matched transitions for which the electronic configurations are  inconsistently reported.} 
\footnotesize
\begin{tabular}{ c c c c c c c c c}
            \hline
            \hline
            \noalign{\smallskip}
            & Ion      &  $\lambda$ (\AA) & $E_{i}$~(eV) & $E_{k}$~(eV) & $J_{i}$ & $J_{k}$ &  Configuration: lower - upper & references \\
            \noalign{\smallskip}
            \hline
            \noalign{\smallskip}
            (a)&\ion{Fe}{I} & 4200.087 & 3.884 & 6.835 & 3 & 3 & 3p$^6$3d$^6$($^5$D)4s4p($^3$P$^\circ$) z$^3$D$^\circ$ - 3p$^6$3d$^7$($^4$F)4d f$^3$F & K07 $^a$ \\
            &\ion{Fe}{I} 	  & 4200.087 & 3.884 & 6.835 & 3 & 3 & ~~~~~~3d$^6$($^5$D)4s4p($^3$P$^\circ$) z$^3$D$^\circ$ - ~~~~~~3d$^7$($^4$F)4d f$^3$F & K14 $^a$ \\
            \noalign{\smallskip}            
            \noalign{\smallskip}
            (b)&\ion{W}{I}  & 4200.028 & 2.037 & 4.988 & 4 & 4 & 5d$^4$6s$^2$ a$^3$G - $^\circ$ & CB $^{b}$\\
            &\ion{W}{I} 	  & 4200.020 & 2.037 & 4.988 & 4 & 4 & 5d$^4$6s$^2$ ~~$^3$G - $^\circ$ & L153 $^{c}$\\
            \noalign{\smallskip}
            \noalign{\smallskip}
            (c)&\ion{Ca}{II}  & 4206.176 & 7.505 & 10.452 & 1/2 & 1/2 & 3p$^6$($^1$S)5p $^2$P$^\circ$ - 3p$^6$($^1$S)8s $^2$S & K99 $^a$ \\
            &\ion{Ca}{II} 	   & 4206.180 & 7.505 & 10.452 & 1/2 & 1/2 & 3p$^6$~~~~~~~5p $^2$P$^\circ$ - 3p$^6$~~~~~~~8s $^2$S & L7323 $^{d}$\\
            \noalign{\smallskip}
            \noalign{\smallskip}
            (d)&\ion{Fe}{I}  & 4202.335 & 2.588 & 5.538 & 3 & 2 & 3p$^6$3d$^6$4s$^2$ b$^3$F~~ - 3p$^6$3d$^6$(a$^3$F~~)4s4p($^3$P$^\circ$) v$^5$D$^\circ$ & K07 $^a$\\
            &\ion{Fe}{I} 	   & 4202.330 & 2.588 & 5.538 & 3 & 2 & ~~~~~~3d$^6$4s$^2$ b$^3$F2 - ~~~~~~3d$^6$( ~$^3$F2)4s4p($^3$P$^\circ$) v$^5$D$^\circ$ & L11631 $^{e}$\\
            \noalign{\smallskip}
            \hline
\end{tabular}
\tablefoot{(d) shows differences in seniority of term and differences in sub-shell. $^a$\citet{klines} $^b$\citet{cbline} $^c$\citet{l153} $^d$\citet{l7323} $^e$\citet{l11631}
}
\label{configdiff}
\end{table*}

BRASS is composed of older versions of NIST and Kurucz transitions, which form the bulk of VALD transitions, so we expect that the majority of BRASS transitions can be cross-matched against atomic transitions in more recent versions of NIST and VALD3. For the non-parametric cross-match of BRASS against VALD3 we successfully cross-matched $\sim$90\% of all BRASS transitions. While the majority of transitions in a given database kept the same configuration nomenclature parsing, we found a number of updated lines for which the nomenclature parsing had changed. The most notable origin of the changes was due to updated Kurucz \ion{Fe}{I} lines, updating from K07 to K14 \citep{klines}, where the reporting of closed sub-shells ceased.

\begin{table}
\centering
\caption{The number of cross-matched transitions, for BRASS cross-matched against all retrieved repositories and line lists, using the parametric and non-parametric cross-matches.}
\begin{tabular}{l r r}
            \hline
            \hline
            \noalign{\smallskip}
            Repository      &  parametric & non-parametric  \\
            \noalign{\smallskip}
            \hline
            \noalign{\smallskip}
            SpectroWeb 	& 36841 & -   \\
            VALD3		& 72178 & 69159  \\
            NIST 	& 9477 & 8943   \\
            Spectr-W$^3$ 	& 1221 & - \\
            TIPbase		& - & ~~~~203~(1123)~{$^a$}  \\
            TOPbase		& - & ~~~~709~(1968)~{$^a$}  \\
            CHIANTI 	& 244 & -   \\
            
            \hline            
\end{tabular}

\tablefoot{$^a$Number of complete and incomplete multiplets. The corresponding number of lines in BRASS is given in parenthesis.}
\label{crossmatchnumbers}
\end{table}

Table~\ref{crossmatchnumbers} shows the number of cross-matched transitions, for BRASS cross-matched against all retrieved repositories and line lists, using the parametric and non-parametric cross-matches. While the number of matches produced by each method are comparable, it is important to note that the parametric method cannot accurately handle large differences in wavelength or energy levels and can thus lead to incorrect cross-matches. On the other hand the non-parametric cross-match depends on the consistency and accuracy of the atomic terms, which can sometimes change over time especially for transitions with significant configuration mixing. This is noted by VALD who opts to use the parametric cross-match when compiling transitions from multiple sources \citep{vamandvald}. An example of such issues with configurations and terms is discussed in Section 5.1.
The added accuracy and reliability of the non-parametric cross-match far outweighs this potential drawback. By using the atomic configuration cross-match BRASS is able to cross-match transitions regardless of the difference between atomic parameters, which can be substantial between older atomic calculations and newer data, and thus provide a robust and thorough compilation of literature transitions. The accuracy of the non-parametric cross-match, compared with the parametric cross-match, is important for such a large homogeneous literature assessment as it ensures that the correct $\log(gf)$ values are compared against each other. The non-parametric method is also an accurate way to compile clean and thorough line lists from scratch, and also to clean existing line lists of potential duplicate transitions, if the atomic configuration information is retained. Finally, the non-parametric cross-match can be used to autonomously compile and work with multiplet tables within our line list. Being able to easily and automatically wield atomic multiplets will enhance the BRASS project's future quality assessment work, by fully exploiting the underlying atomic physics investigations of systematic $\log(gf)$ correlations of spectral lines.

\subsection{Parametric cross-match}
Care was taken to ensure that all repository wavelength values were converted to air wavelengths and all energies and wavelengths used the same units. We employed the same air to vacuum wavelength conversion as VALD3\footnote{\url{http://www.astro.uu.se/valdwiki/Air-to-vacuum\%20conversion}}. The conversion from $\lambda_{vac}$ to $\lambda_{air}$, taken from \citet{airconversion}, applies to dry air at 1 atm pressure and 15$^\circ$C with 0.045\% CO$_{2}$ by volume, and is as follows:	
\begin{equation}
\begin{gathered}
      \lambda_{air}  =  \frac{\lambda_{vac}}{n}
\end{gathered}
\end{equation}
\begin{equation}
\begin{gathered}
      n  = {1 + 0.0000834254 + \frac{0.02406147}{130-s^2} + \frac{0.00015998}{38.9-s^2}}
\end{gathered}
\end{equation}
where the refractive index, $n$, is taken from \citet{airvacuum}, $s$~=~10$^{4} / \lambda_{vac}$, and $\lambda_{vac}$ is in \aa ngstr\"{o}ms.

For our parametric cross-match work we use thresholds on $\lambda$ and energies as well as requiring cross-matched transitions to have the same upper and lower total angular momentum \textit{J}-values. While $\lambda$ is inversely proportional to $E_{k} - E_{i}$, it is important to note that an individual database transition may be compiled from several sources, both experimental and theoretical, which can lead to discrepancies for transitions with experimentally measured wavelengths and theoretical energy levels. Due to this we have chosen to treat our wavelength and energy level thresholds independently. Using our non-parametric cross-match we were able to determine wavelength and energy level thresholds according to the $\Delta\lambda$ and $\Delta E$ distributions, as shown later in the paper by Figure~\ref{part5t} and Figure~\ref{part67t}. According to these distributions we chose a wavelength threshold of $\lambda~\pm$~0.1\AA~and energy thresholds of $E~\pm$~0.0005~eV.

These thresholds reproduced $\sim$95\% of the non-parametric cross-matches for both VALD and NIST. The parametric cross-match proves to be a useful cross-match compromise when configuration information is lacking or difficult to use. However it is important to re-iterate that there is still no fundamental guarantee that two automatically matched transitions are meant to be the same physical transition, especially in ions with complicated electronic structures such as Fe-group elements. Table~\ref{incorrectmatch} shows an example of one correctly cross-matched transition pair and two incorrectly matched transition pairs, using different energy thresholds of $E~\pm$~0.1eV and $E~\pm$~0.0005~eV. The correctly matched pair was found by both the parametric and non-parametric cross-matches, whereas only the parametric method incorrectly matched the remaining pairs of transitions.

\begin{table*}
\centering
\caption{Examples of cross-matched transitions for which the parametric method is insufficient to accurately cross-match transitions. } 
\footnotesize
\begin{tabular}{ c c c c c c c c c}
            \hline
            \hline
            \noalign{\smallskip}
            & Ion      &  $\lambda$ (\AA) & $E_{i}$~(eV) & $E_{k}$~(eV) & $J_{i}$ & $J_{k}$ &  Configuration: lower - upper & references \\
            \noalign{\smallskip}
            \hline
            \noalign{\smallskip}
           (a) &\ion{Mn}{II}	& 4639.152 & 10.774 & 13.446 & 3 & 2 & 3d$^4$($^5$D)4s4p($^3$P$^\circ$) w$^5$P$^\circ$ - 3d$^5$($^6$S)7s $^5$S &K09 $^a$\\
           &\ion{Mn}{II} 	& 4639.160 & 10.774 & 13.446 & 3 & 2 & 3d$^4$($^5$D)4s4p($^3$P$^\circ$) w$^5$P$^\circ$ - 3d$^5$($^6$S)7s $^5$S & L562 $^{b}$\\
            \noalign{\smallskip}
            \noalign{\smallskip}
            ~(b)$^*$&\ion{Cr}{I}  & 4244.770 & 3.890 & 6.810 & 4 & 5 & ~~~~~~~~~~~~~~~~~~~3d$^5$($^4$F)4s a$^5$F~~ - 3d$^5$($^4$G)5p v$^3$H$^\circ$ & K10 $^a$\\
            &\ion{Cr}{I} 	   & 4244.340 & 3.857 & 6.777 & 4 & 5 & ~~~~3d$^4$($^5$D)4s4p($^3$P$^\circ$) z$^5$F$^\circ$ ~- ~~~~~~~~~~~~~~~~~~e$^5$F & L808 $^{c}$ \\
            \noalign{\smallskip}
            \noalign{\smallskip}
            ~~(c)$^{**}$&\ion{Fe}{II}  & 6207.273 & 11.051 & 13.048 & 7/2 & 5/2 & ~~~~~~~~3d$^6$($^5$D)5p $^6$P$^\circ$ - 3d$^6$($^5$D)5d $^6$S & K13 $^a$ \\
            &\ion{Fe}{II} 	   & 6207.342 & 11.051 & 13.048 & 7/2 & 5/2 & 3d$^5$($^4$P)4s4p($^3$P)  $^6$P$^\circ$ - 3d$^6$($^5$D)5d $^6$G  & RU $^d$\\
            \noalign{\smallskip}
            \hline
\end{tabular}
\tablefoot{$^*$Cross-matched using $\Delta E~=~\pm$~0.1~eV $^{**}$Cross-matched using $\Delta E~=~\pm$~0.0005~eV  $^a$\citet{klines} $^b$\citet{l562} $^c$\citet{l808} $^d$\citet{Raassen}
}
\label{incorrectmatch}
\end{table*}

\subsection{TIPbase cross-matches}

Contrary to the previous databases, which provide experimentally measured wavelengths and a mix of experimental and theoretical oscillator strengths, the iron project database, TIPbase, \citep{tipbase} and the opacity project atomic database, TOPbase, \cite{topbase} offer pure theoretical atomic data obtained using ab initio quantum mechanical calculations. 
The goal of the long-term opacity project was to provide reliable atomic data to estimate stellar envelope opacities from photo-ionisation cross-sections and oscillator strengths.
Due to the complexity of the task that required model atoms, fine structure was barely considered in the ab initio quantum calculations. 
As a consequence, atomic lines for TIPbase and TOPbase are given as multiplets, without fine-structure \textit{J}-values, with average wavelengths and weighted average oscillator strengths rather than components with individual wavelengths and oscillator strengths.

TIPbase focuses on the calculation of
ab initio electron excitation cross-sections for iron and its
ions \citep{tipbase}. Unfortunately, the data concerning
\ion{Fe}{i} and \ion{Fe}{ii} are not available via the VAMDC or the TIPbase data
server\footnote{\url{http://cdsweb.u-strasbg.fr/tipbase/home.html}}. To
retrieve them, we found access to the raw outputs through the Nahar OSU
radiative atomic
database\footnote{\url{http://www.astronomy.ohio-state.edu/\~nahar/nahar_radiativeatomicdata}}
from which we extracted the required data, and formatted data to be consistent with our current work.

We used a non-parametric approach to cross-match the TIPbase
transitions with BRASS lines based on the identification of spectral
term and seniority index. This is normally sufficient to unequivocally
identify lines in BRASS belonging to a given TIPbase transition. No
wavelength restriction has been used because the difference between
theoretical and experimental energy levels can reach several tenths of
eV, leading to difference in the wavelengths of up to thousands of \aa ngstr\"{o}ms.
For instance, the weighted average wavelength of the multiplet
y$^5$F$^\circ$ -- f$^5$D reconstructed from the components in BRASS is
6134~\AA\ whereas it is given at 4559~\AA\ in TIPbase, a difference
of 1575~\AA~(equivalent to a difference of 0.70 eV).

Transitions between two different multiplicity systems do not occur in
TIPbase data and thus the identification of the lines only occurs within a given multiplicity. In spin-orbit coupling, multiplicities of one, three, five, and seven exist for
\ion{Fe}{i} and of two, four, six, and eight for \ion{Fe}{ii}. The cross-matches
with BRASS occur for lines within one, three, and five multiplicity systems for
\ion{Fe}{i} and two, four, and six for \ion{Fe}{ii}. We identified
121 \ion{Fe}{i} and 74 \ion{Fe}{ii} TIPbase multiplets corresponding to
700 \ion{Fe}{i} and 423 \ion{Fe}{ii} BRASS lines. Among them 47
\ion{Fe}{i} and 39 \ion{Fe}{ii} TIPbase multiplets are incomplete,
meaning that one or several BRASS components are missing to complete a
TIPbase multiplet due to the wavelength restrictions on the BRASS atomic line list. Table~\ref{tab:xmatch_tip_brass} shows a detailed breakdown of the cross-match Fe multiplicity.

For TIPbase we have attempted to calculate the oscillator strengths of the
components within a TIPbase multiplet. Using the relative line strengths
within a normal multiplet, shown in Tables 4.7, 4.8 and 4.9 of the
Allen's Astrophysical Quantitities \citet{allenastro}), we calculated the
oscillator strengths of the components of \ion{Fe}{i} multiplet 15
(a$^5$F -- z$^5$D$^\circ$) from the TIPbase oscillator strength of the
multiplet (Table~\ref{tab:comp_tip_brass}). The strongest components
are the principal lines (labelled $x$) the weakest ones are the
satellite lines (labelled $y$ and $z$). The differences between
TIPbase and BRASS oscillator strengths for this multiplet are within a
factor of two.
 Such detailed comparisons were never previously done as "no attempt is
made to include fine structure effects in our calculated energy
levels, oscillator strengths and photoionization cross
sections" \citep{saweyberrington}. Comparisons were only made on the summed $gf$-values, that is multiplet by multiplet.
For instance, \citet{bautista1997} did a comparison of
multiplet $gf$-values with experiments from \citet{nave1994} and \citet{FMW}, as given in his Table 5. He quoted that the differences in
multiplet gf-values, when compared with the experimental ones, can reach 40\%
in the worst cases. In the case of the multiplet n$^{\circ}$15 (a$^5$F
-- z$^5$D$^\circ$), we find a 28\% difference between TIPbase and BRASS.
Corrections using the observed energy levels  only decrease the percentage to
27\%. For individual components, the deviations can reach up to 38\%, even for
the main component. Such deviations are expected since the TIPbase
oscillator strengths were calculated in pure LS coupling. These
deviations from LS coupling are mainly due to the electronic
configuration mixing processes and possibly relativistic effects in the radial part of the dipole operator. 

\begin{table}
\centering
\caption{\ion{Fe}{i} and \ion{Fe}{ii} cross-matches of TIPbase with
BRASS data.} \label{tab:xmatch_tip_brass}
    \begin{tabular}{rcc}
\hline
\hline
\noalign{\smallskip}
Multiplicity & TIPbase & BRASS \\
\hline
\noalign{\smallskip}
         \ion{Fe}{i} & \# multiplets & \# lines \\
\hline
\noalign{\smallskip}
Complete & 74 & 404\\
         1 & 20&20\\
         3 & 40 & 248\\
         5 & 14 & 136\\
\noalign{\smallskip}
Incomplete& 47 & 296\\
         3&31&155\\
         5&16&141\\
\hline
\noalign{\smallskip}
         Total & 129 & 700 \\
\hline
\noalign{\smallskip}
\hline
\noalign{\smallskip}
\ion{Fe}{ii} & \# multiplets & \# lines \\
 \hline
\noalign{\smallskip}
          Complete & 35 & 227 \\
          2&20&85\\
          4&14&139\\
          6&1&3\\
\noalign{\smallskip}
 Incomplete & 39 & 196\\
          2&13&27\\
          4&26&169\\
          \hline
\noalign{\smallskip}
          Total & 74 & 423 \\
 \hline

 \end{tabular}
 \end{table}

 \begin{table*}
 \centering
 \caption{Comparison of the estimated TIPbase $\log(gf)$ for the
 components of the \ion{Fe}{i} multiplet n$^\circ$15 (a$^5$F --
 z$^5$D$^\circ$) with the BRASS values. The relative deviations are calculated on the $gf$-values.} \label{tab:comp_tip_brass}
 \begin{tabular}{cccccccrr}
\hline
\hline 
\noalign{\smallskip}
  $\lambda$  & $J_\mathrm{i}$ &
 $J_\mathrm{k}$ & Component & fraction & $\log{gf}_\mathrm{TIP}$ &
 $\log{gf}_\mathrm{BRASS}$ & $\Delta \log{gf}$ & Relative \\ ~(\AA)
 &                  &                 &           &
 &                         &                           &
 TIP-BRASS     & deviation (\%) \\ \hline \\ Components \\

 5269.5370 & 5 & 4 & $x_1$ & 0.31428 & $-1.117$ & $-1.324$ & $0.207$ &
 38 \\ 5328.0386 & 4 & 3 & $x_2$ & 0.04286 & $-1.283$ & $-1.466$ &
 $0.183$ & 34 \\ 5371.4893 & 3 & 2 & $x_3$ & 0.00114 & $-1.477$ &
 $-1.645$ & $0.168$ & 32 \\ 5397.1279 & 4 & 4 & $y_1$ & 0.21429 &
 $-1.982$ & $-1.991$ & $0.008$ &  2 \\ 5405.7746 & 2 & 1 & $x_4$ &
 0.06000 & $-1.711$ & $-1.849$ & $0.138$ & 27 \\ 5429.6964 & 3 & 3 &
 $y_2$ & 0.00571 & $-1.836$ & $-1.879$ & $0.043$ &  9 \\ 5434.5235 & 1 &
 0 & $x_5$ & 0.13714 & $-2.012$ & $-2.121$ & $0.109$ & 22 \\ 5446.9164 &
 2 & 2 & $y_3$ & 0.05714 & $-1.857$ & $-1.914$ & $0.057$ & 12 \\
 5455.6091 & 1 & 1 & $y_4$ & 0.00571 & $-2.012$ & $-2.093$ & $0.081$ &
 17 \\ 5497.5160 & 1 & 2 & $z_3$ & 0.08000 & $-2.857$ & $-2.845$ &
 $-0.012$&  3 \\ 5501.4649 & 3 & 4 & $z_1$ & 0.04000 & $-3.158$ &
 $-3.046$ & $-0.112$& 29 \\ 5506.7787 & 2 & 3 & $z_2$ & 0.04000 &
 $-2.857$ & $-2.795$ & $-0.062$& 15 \\ \\ TIPbase multiplet \\

 5464.2500$^a$ &17 & 12 & & 1.00 & $-0.615$ & $-0.760$ & $0.145$ & 28 \\
 \hline
 \end{tabular}
 \tablefoot{$^a$TIPbase wavelength}
 \end{table*}

\subsection{TOPbase cross-matches}

In the TOPbase data retrieved from VAMDC, there are entries for 18 elements\footnote{H, He, Li, Be, B, C, N, O, F, Ne, Na, Mg, Al, Si, S, Ar, Ca and \ion{Fe}{iii}} over 3 ionization degrees. 
The cross-matching process has been performed between TOPbase and BRASS, in the same manner as TIPbase, and examples of such matches are presented in Table~\ref{topbasetable}.
For a given ion and multiplet in TOPbase our cross-match retrieves all the associated individual fine-structure transitions in BRASS. In the range 4200-6800~\AA, we have identified 136 singlets\footnote{\ion{Be}{i}, \ion{Be}{iii},\ion{C}{i, }\ion{C}{iii}, \ion{N}{ii}, \ion{O}{iii},\ion{F}{ii}, \ion{Mg}{i}, \ion{Al}{ii}, \ion{Si}{i}, \ion{S}{i}, \ion{Ca}{i} and \ion{Fe}{iii}} and 573 multiplets\footnote{corresponding to 32 different species} with 1968 BRASS transitions.
It is important to re-emphasise the inaccuracy of cross-matching either TIPbase or TOPbase transitions using a parametric cross-match. The theoretical wavelengths are often inaccurate due to (i) differences in theoretical and experimental energy levels and (ii) the neglect of fine structure. Examples of this case are (i) the match for the \ion{Mg}{i} singlet line in Table~\ref{topbasetable} where the TOPbase wavelength is 40~\AA\ larger than the corresponding BRASS line; and (ii) the \ion{Ca}{ii} multiplet in Table~\ref{topbasetable} where the mean wavelength is almost 100~\AA\ larger than the two associated BRASS transitions. In the context of our future atomic data quality assessment, the multiplets are first decompounded into components using the LS intensities calculated with the Wigner $6j$ coefficients, as described in Section 3.3 and shown in Table~\ref{tab:comp_tip_brass}, and then scaled to the database $gf$ multiplet.

\begin{table*}
\centering
\caption{Examples of singlet~-~(a) and multiplet~-~(b) cross-matches between TOPbase and BRASS. The first transition in each set of matches corresponds to the TOPbase singlet or multiplet transition and all subsequent transitions are fine structure transitions present in BRASS. }
 \footnotesize
 \begin{tabular}{cccccccccrr}
\hline\hline
  & Ion  & $\lambda$ (\AA) & $E_{\mathrm{i}}$ (eV) &  $E_{\mathrm{k}}$ (eV) & $J_{\mathrm{i}}$ & $J_{\mathrm{k}}$ & Configuration & Term&$\log{gf}$ & $\Delta \log{gf}$\\
\hline \\
	(a) & \ion{O}{iii} & 5291.416   & 39.156   & 41.508   & 0    & 1   &  3p - 3d                                                        & $^1$S - $^1$P$^{\circ}$ & $-0.326$ &$-0.033$ \\
	 & \ion{O}{iii} & 5268.301   & 38.907   & 41.260   & 0    & 1   & 2s$^2$2p($^2$P$^{\circ}$)3p - 2s$^2$2p($^2$P$^{\circ}$)3d & $^1$S - $^1$P$^{\circ}$ & $-0.359$ &         \\
\noalign{\smallskip}
\noalign{\smallskip}
     (a) & \ion{Mg}{i} &  4391.381   & 4.422    & 7.255    & 1    & 2   &  3p  -     6d                				                     & $^1$P$^{\circ}$ - $^1$D &  $-0.583$ &$0.000$ \\
     & \ion{Mg}{i} &  4351.906   & 4.346    & 7.194    & 1    & 2   & 3s3p  - 3s6d         				                     & $^1$P$^{\circ}$ - $^1$D &  $-0.583$ &        \\
\noalign{\smallskip}
\noalign{\smallskip}
     (a) & \ion{Si}{i} &  6361.851   & 5.131    & 7.086    & 1    & 1   &  4s  -     5p                 				                     & $^1$P$^{\circ}$ - $^1$P & $-2.712$ &$ 0.895$ \\
     & \ion{Si}{i} &  6331.956   & 5.082    & 7.040    & 1    & 1   &  3s$^2$3p4s - 3s$^2$3p5p  				                 & $^1$P$^{\circ}$ - $^1$P & $-1.817$ &         \\
\noalign{\smallskip}
\hline
\noalign{\smallskip}     
  (b)   & \ion{Al}{i} &  6752.439   & 3.110    & 4.952    & 1/2 & 5/2 	&  4s - 5p 									                     & $^2$S - $^2$P$^{\circ}$ & $-1.396$ &$0.003$ \\
     & \ion{Al}{i} &  6696.023   & 3.143    & 4.994    & 1/2 & 3/2   & 3s$^2$4s - 3s$^2$5p                                      & $^2$S - $^2$P$^{\circ}$ & $-1.569$ &        \\
	 & \ion{Al}{i} &  6698.673   & 3.143    & 4.993    & 1/2 & 1/2   & 3s$^2$4s - 3s$^2$5p                                      & $^2$S - $^2$P$^{\circ}$ & $-1.870$ &        \\
\noalign{\smallskip}
\noalign{\smallskip}
  (b)   & \ion{S}{ii} &  5012.721   & 14.012   & 16.494   & 11/2 & 11/2 & 4s - 4p                                                        & $^4$P - $^4$P$^{\circ}$ & $ 0.614$ &$0.007$ \\
     & \ion{S}{ii} &  4924.110   & 13.618   & 16.135   & 3/2  & 5/2  & 3s$^2$3p$^2$($^3$P)4s - 3s$^2$3p$^2$($^3$P)4p            & $^4$P - $^4$P$^{\circ}$ & $-0.059$ &        \\ 
     & \ion{S}{ii} &  4925.343   & 13.584   & 16.101   & 1/2  & 3/2  & 3s$^2$3p$^2$($^3$P)4s - 3s$^2$3p$^2$($^3$P)4p            & $^4$P - $^4$P$^{\circ}$ & $-0.235$ &        \\ 
     & \ion{S}{ii} &  5009.567   & 13.618   & 16.092   & 3/2  & 1/2  & 3s$^2$3p$^2$($^3$P)4s - 3s$^2$3p$^2$($^3$P)4p            & $^4$P - $^4$P$^{\circ}$ & $-0.094$ &        \\ 
     & \ion{S}{ii} &  5032.434   & 13.672   & 16.135   & 5/2  & 5/2  & 3s$^2$3p$^2$($^3$P)4s - 3s$^2$3p$^2$($^3$P)4p            & $^4$P - $^4$P$^{\circ}$ & $ 0.282$ &        \\ 
\noalign{\smallskip}
\noalign{\smallskip}
	(b) & \ion{Ca}{ii} & 4300.578   & 7.459    & 10.352   & 5/2  & 1/2  & 5p - 8s                                                        & $^2$P$^{\circ}$ - $^2$S & $-1.128$ &$-0.605$ \\
     & \ion{Ca}{ii} & 4206.176   & 7.505    & 10.452   & 1/2  & 1/2  & 3p$^6$($^1$S)5p - 3p$^6$($^1$S)8s                        & $^2$P$^{\circ}$ - $^2$S & $-2.210$ &         \\	 
     & \ion{Ca}{ii} & 4220.071   & 7.515    & 10.452   & 3/2  & 1/2  & 3p$^6$($^1$S)5p - 3p$^6$($^1$S)8s                        & $^2$P$^{\circ}$ - $^2$S & $-1.910$ &         \\
\noalign{\smallskip}
\noalign{\smallskip}
\hline
\end{tabular}
\label{topbasetable}
\end{table*}

\section{Differences between repository transitions}

\subsection{Differences with our reference list}

\begin{figure*}
\centering
\includegraphics[width=16cm]{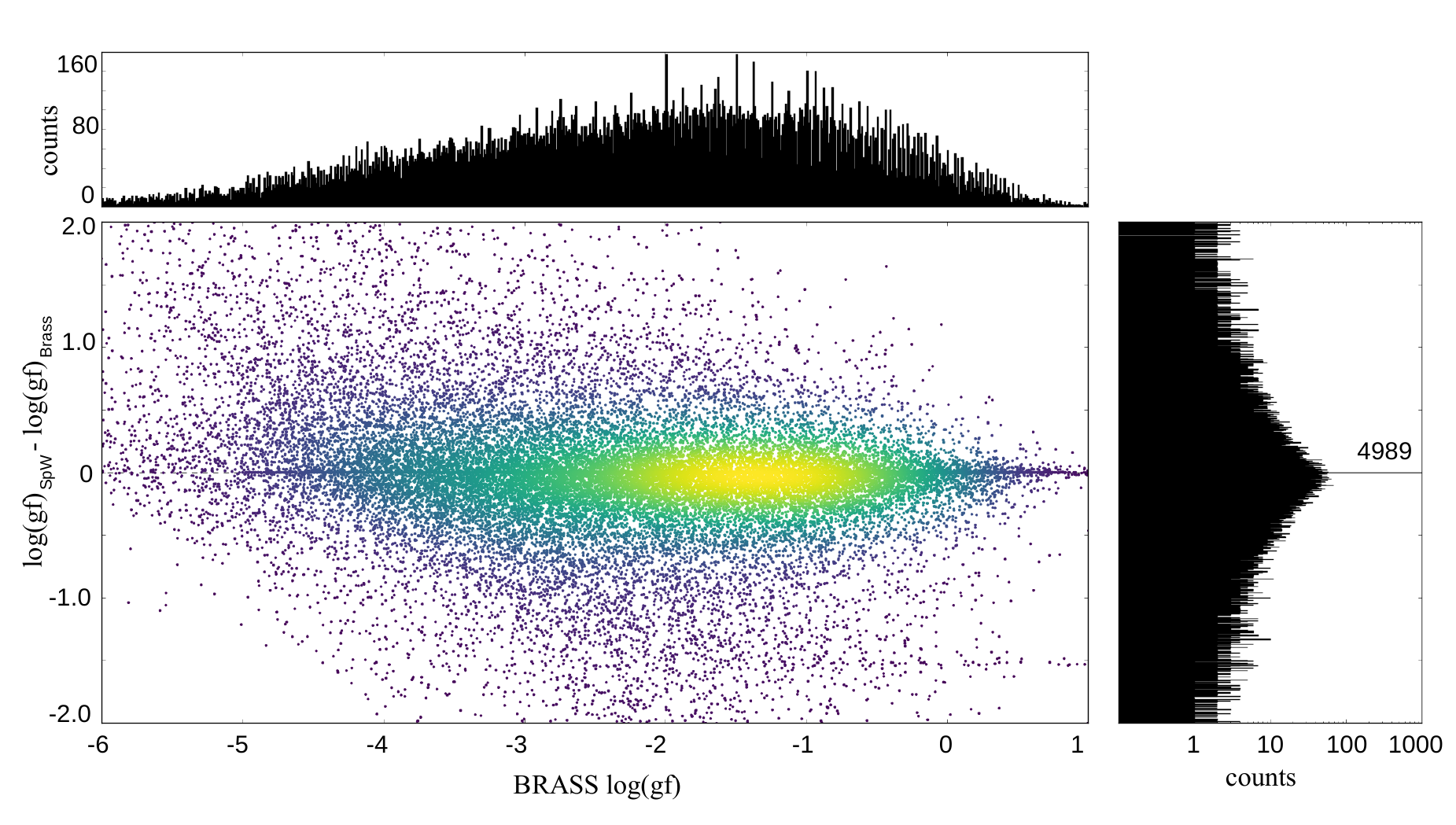}
\caption{ Distribution of $\log(gf)$ vs $\Delta \log(gf)$ (SpectroWeb - BRASS) for the SpectroWeb(2008) and BRASS(2012) parametric cross-match. For clarity and consistency a small number of transitions with $|\Delta \log(gf)|$~>~2 have been excluded. In some extreme cases these $\Delta \log(gf)$ values can be as large as $\Delta \log(gf)~\pm$4~dex. Compared to the other non-parametric cross-matches in Figures~\ref{part2} and \ref{part4} we see a substantial increase in scatter caused mainly by updates to the literature, but also due to the lack of \textit{J}-values to further constrain the parametric cross-match. 
              }
\label{part1}
\end{figure*}

\begin{figure*}[!t]
\centering
\includegraphics[width=16cm]{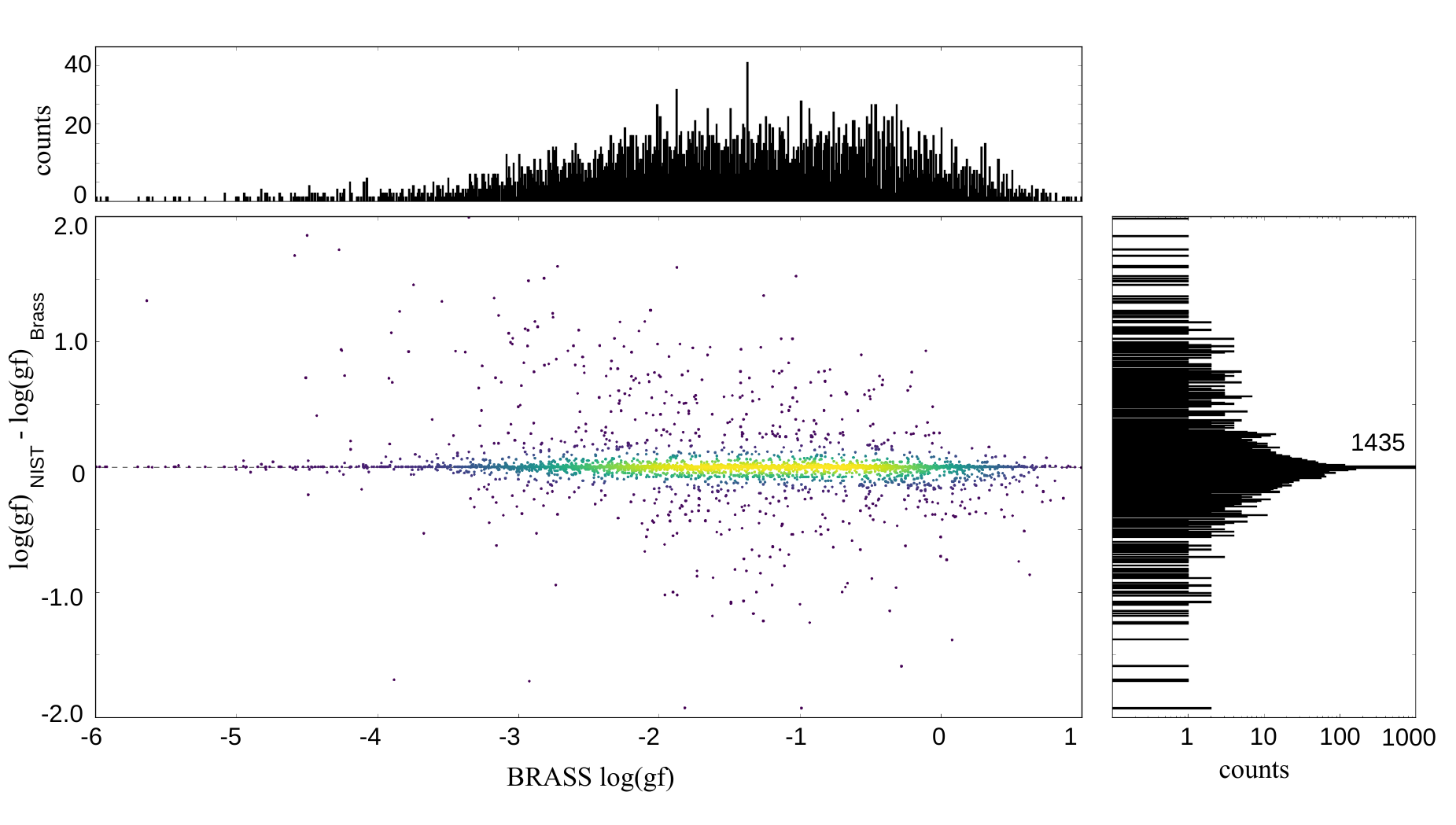}
\caption{ Distribution of $\log(gf)$ vs $\Delta \log(gf)$ (NIST - BRASS) for the NIST(2016) and BRASS(2012) non-parametric cross-match. For clarity and consistency a small number of transitions with $|\Delta \log(gf)|$~>~2 have been excluded. In some extreme cases these $\Delta \log(gf)$ values can be as large as $\Delta \log(gf)~\pm$2~dex. When compared with Figures~\ref{part1} and \ref{part4} we find that NIST is weighted more towards stronger transitions than SpectroWeb or VALD, no doubt due to the experimental provenience of NIST transitions.
              }
\label{part2}
\end{figure*}

\begin{figure*}[!t]
\centering
\includegraphics[width=16cm]{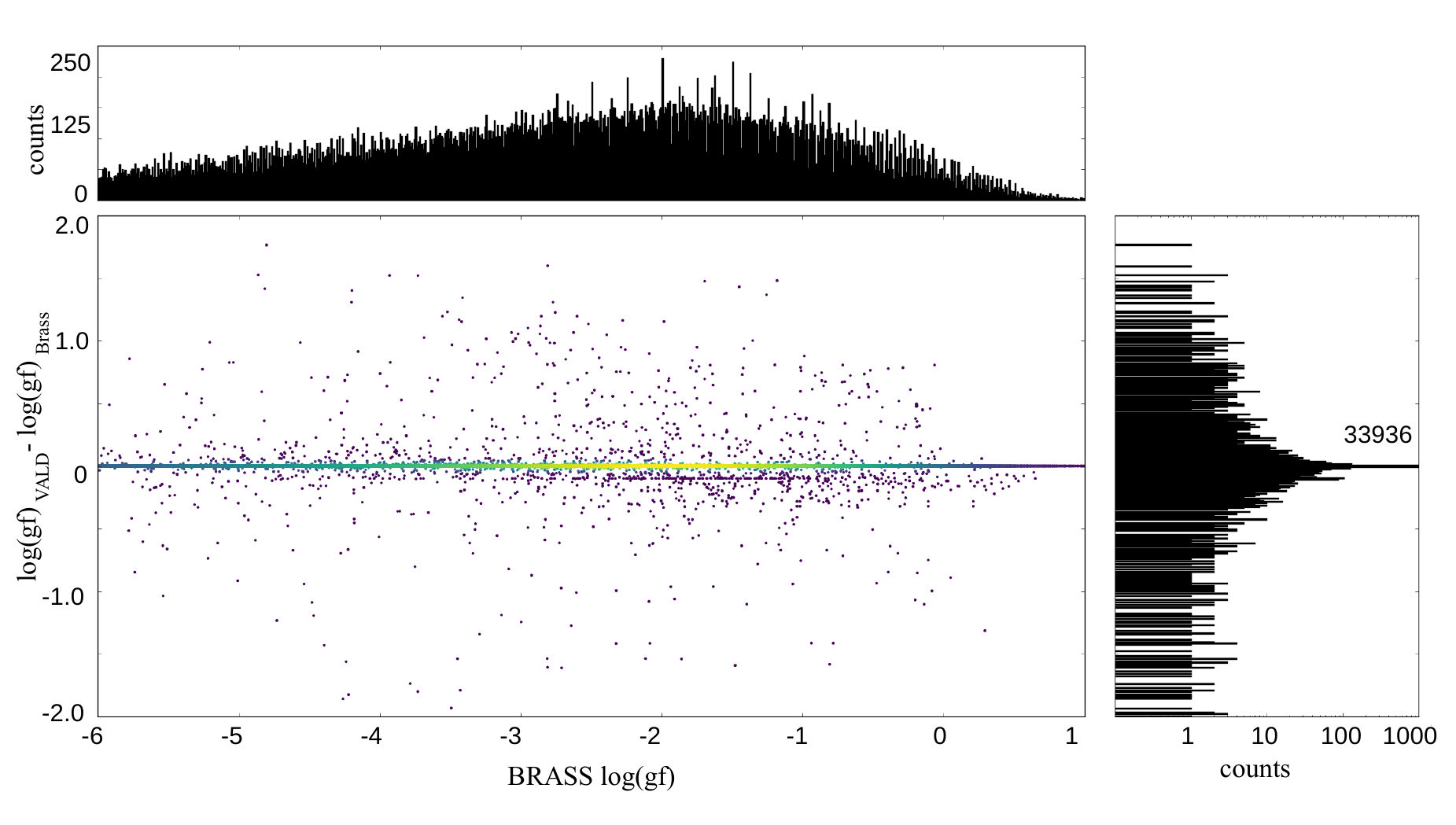}
\caption{ Distribution of $\log(gf)$ vs $\Delta \log(gf)$ (VALD3 - BRASS) for the VALD3(2016) and BRASS(2012) non-parametric cross-match. For clarity and consistency a small number of transitions with $|\Delta \log(gf)|$~>~2 have been excluded. In some extreme cases these $\Delta \log(gf)$ values can be as large as $\Delta \log(gf)~\pm$~4~dex.  We see a secondary peak in $\Delta \log(gf)$ values at $\Delta \log(gf)$~=~-0.1~dex. This is due to differences in \ion{Fe}{I} lines from \cite{MRW} and \cite{FMW}: both use the $\log(gf)$ values of \cite{bridges} however \cite{FMW} apply a systematic shift to the $\log(gf)$ values of -0.1~dex.
              }
\label{part4}
\end{figure*}

\begin{figure*}[!t]
\centering
\includegraphics[width=16cm]{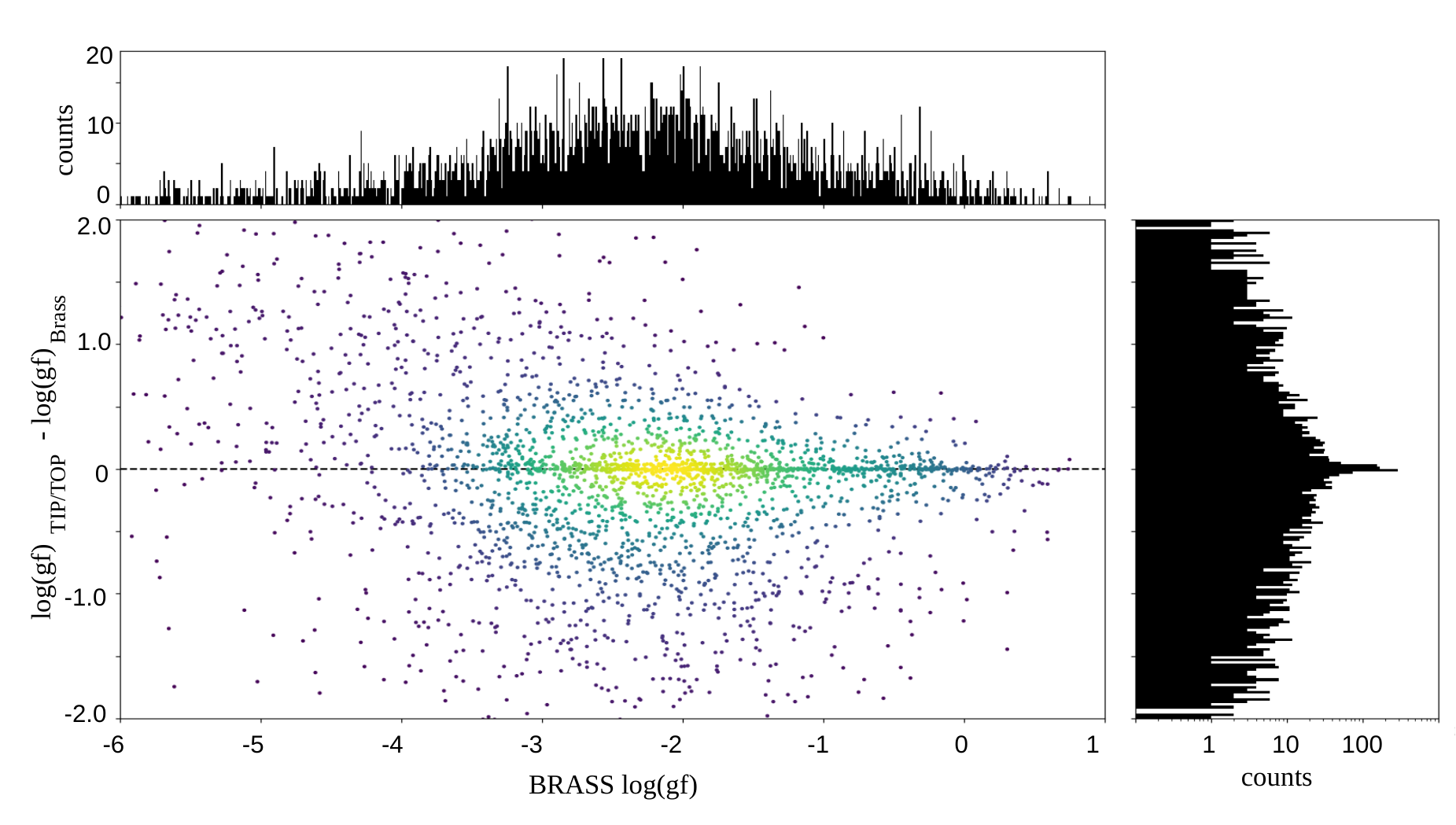}
\caption{ Distribution of $\log(gf)$ vs $\Delta \log(gf)$ (TIPTOP - BRASS) for the TIPTOP and BRASS(2012) cross-matches. For clarity and consistency a number of transitions with $|\Delta \log(gf)|$~>~2 have been excluded. In a number of cases these $\Delta \log(gf)$ values are as large as $\Delta \log(gf)~\pm$~4~dex.
              }
\label{part4t}
\end{figure*}

SpectroWeb was cross-matched against BRASS using the parametric method while NIST and VALD were cross-matched against BRASS using the non-parametric method. TIPbase and TOPbase were also cross-matched against BRASS using non-parametric methods and have been grouped together as TIPTOP in the following analysis. The BRASS, VALD and NIST lists have all been cleaned of duplicate transitions, discussed in Section 6, prior to all cross-matches. Transition $\log(gf)$ values play an important role in the spectroscopic analysis of stars, and are related to the equivalent width of a spectral line $W_{\lambda}$ according to the following equation that defines the curve of growth \citep{gray1992}:

\begin{equation}
\begin{gathered}
      \log \left (\frac{W_{\lambda}}{\lambda} \right ) = \log(gf\lambda) + \log(A_{el})  - \frac{5040}{T_e}E_{i} +\log(C)
\end{gathered}
\end{equation}
where $\lambda$ is the transition wavelength, $A_{el}$ is the elemental abundance of the spectral line, $T_e$ is the excitation temperature, $E_i$ is the lower energy level, and $C$ contains the remaining terms such as the Saha population factor and the continuum opacity, constants for a given star and a given ion. One typically measures the equivalent width $W_{\lambda}$ of a spectral line and uses the atomic data and known stellar parameters to derive an elemental abundance $A_{el}$ for the stellar line in question. For lines that fall on the linear part of the curve of growth this implies that any change in the adopted $\log(gf)$ value will have an inversely proportional impact on the derived abundance for that stellar line. With this relationship in mind, Figures~\ref{part1}-\ref{part4t} show the cross-matched $\log(gf)_{Database}~-~\log(gf)_{BRASS}$ values plotted against BRASS $\log(gf)$, for SpectroWeb, NIST, VALD, and TIPTOP.

Figures~\ref{part1}-\ref{part4t} all show a wedge-like structure where $\Delta \log(gf)$ appears to increase with decreasing $\log(gf)$. We attribute this relationship to the decrease in accuracy, within multiplets, going from the leading line to subordinate transitions. In Figure~\ref{part1} we observe a cut-off region located at the bottom-left of the plot. This cut-off is caused by the SpectroWeb line list which typically does not contain transitions below $\log(gf)$~=~$-$6 and thus introduces a lower limit on the $\Delta \log(gf)$ values as the BRASS $\log(gf)$ approaches $\log(gf)$~=~$-$6. We also find a small number of transitions that appear to be shifted by roughly $\Delta \log(gf)~=~-1.5$. These cross-matched lines belong to the \ion{Ni}{II} Kurucz 2003 lines, present in the BRASS line list, and the much older Kurucz 1995 \ion{Ni}{II} lines, present in the SpectroWeb line list. Figures~\ref{part2} and~\ref{part4} also contain similar cut-offs located at $\log(gf)~=~-10$. These cut-offs are not shown in the plots as these transitions are typically too weak to appear in stellar spectra and cannot be assessed as part of the BRASS project. The cut-off at $\log(gf)~=~-10$ is most likely caused by imposed limits on atomic calculations of $\log(gf)$ values as such weak transitions are not currently considered to be of interest. At small values of $\Delta \log(gf)$ we do not observe any systematic correlations other than precision limits of $\log(gf)$ and $\Delta \log(gf)$ values.
We find a much larger scatter in $\Delta \log(gf)$ for the SpectroWeb cross-match, shown in Figure~\ref{part1}, than for the NIST cross-match and VALD cross-match, shown in Figure~\ref{part2} and Figure~\ref{part4}, respectively. This is mainly caused by updates to literature values between the SpectroWeb and BRASS list compilations, but a very small number are caused by incorrectly matched transitions due to lack of \textit{J}-values when using the parametric cross-match, which can lead to the comparison of two entirely different $\log(gf)$ values. We also find a larger scatter for the TIPTOP cross-match that is caused by significant updates to theoretical atomic calculations over the last two decades.
A secondary peak in $\Delta \log(gf)$ is visible in Figure~\ref{part4} for VALD3 at $\Delta \log(gf)$~=~-0.1~dex. This is caused by differences in adopted $\log(gf)$ values for a number of \ion{Fe}{I} lines in the BRASS atomic line list and VALD. The BRASS atomic line list has adopted the $\log(gf)$ values from the work of \citet{MRW} whereas the VALD transitions adopt the $\log(gf)$ values from the work of \citet{FMW}. Both works use the experimental $\log(gf)$ values of \citet{bridges}, however Fuhr, Martin \& Weise normalised the Bridges \& Kornblith $\log(gf)$ values of $\log(gf)~\leq$~0.75 by $-0.1$~dex, thus leading to the secondary peak in $\Delta \log(gf)$. 

For the SpectroWeb, NIST, and VALD cross-matches we see a clear peak in $\Delta \log(gf)$~=~0, and especially so for VALD. This is unsurprising as the BRASS line list was created using NIST transitions, and using Kurucz transitions which make up a substantial portion of the VALD lines. The result is that BRASS is essentially an older version of both NIST and VALD combined, and so contains a large number of identical lines, to both NIST and VALD, with $\Delta \log(gf)$~=~0 and a number of updated lines with $\Delta \log(gf)~\neq$~0. For the TIPTOP cross-match this peak is smaller but still present. In addition to the $\Delta \log(gf)$~=~0 we find substantial scatter in $\Delta \log(gf)$, typically in the range of $\Delta \log(gf)~\pm$~0.5~dex, but with a number of transitions reaching $\Delta \log(gf)~\pm$~4~dex.

\begin{figure*}[!t]
\centering
\includegraphics[width=18.5cm]{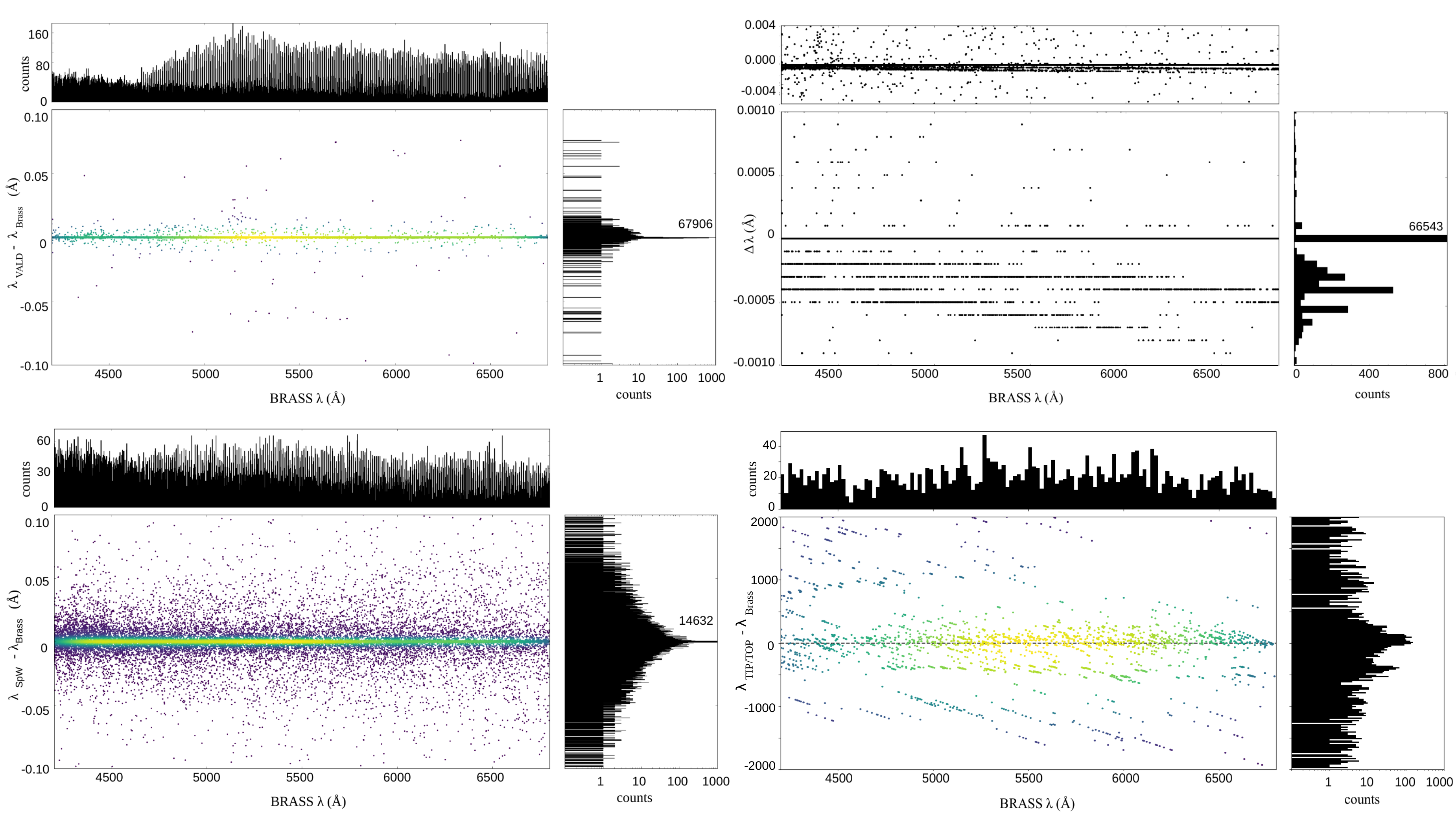}
\caption{ (Top left) $\lambda$ vs $\Delta\lambda$ for VALD3 vs BRASS. (Top right) Small-scale correlations of $\lambda$ vs $\Delta\lambda$ for VALD3 vs BRASS. (Bottom Left) $\lambda$ vs $\Delta\lambda$ for SpectroWeb vs BRASS, which are cross-matched using a $\lambda$ tolerance of    $\pm~0.1~\AA$. (Bottom right) $\lambda$ vs $\Delta\lambda$ for TIPTOP vs BRASS. }
\label{part5t}
\end{figure*}

In order to understand the origin of the scatter of $\log(gf)$ values it is important to note that $gf$ is dependent on the energy levels, $E_{i}$ and $E_{k}$, and the transition line strength $S_{ik}$. The following relationship is true for electric dipole transitions \citep{allenastro}: 

\begin{equation}
\begin{gathered}
      g_i f_{ik} = \frac{8 \pi^2 m_e c }{3 h e^2 \lambda}S_{ik}
\end{gathered}
\end{equation}
where $E_i$ and $E_k$ are written in terms of the transition wavelength $\lambda$, $m_e$ is the electron mass, $c$ is the speed of light, $h$ is the Planck constant, and $e$ is the charge of the electron. Scatter in the $\log(gf)$ values originates from either differences between the wavelengths or the $S_{ik}$ values of the two cross-matched transitions. Figure~\ref{part5t} shows that the differences in wavelength for VALD, NIST, and SpectroWeb are very small and cannot cause the observed scatter in $\log(gf)$ values. For TIPTOP the significant differences in wavelength do contribute towards the large scatter in $\log(gf)$ values. The main cause for the large changes in $\log(gf)$ values is due to the treatment of configuration mixing when calculating $S_{ik}$ values. \citet{fe2calc} note that calculated line strengths are sensitive to the mixing of configuration state functions which can lead to oscillator strength differences of upto a few orders of magnitude when it is included in theoretical quantum mechanical calculations. The majority of VALD transitions, and almost half of NIST transitions, rely on theoretical oscillator strength values, and so changes in the treatment of configuration mixing are likely causes for the large $\Delta\log(gf)$ values we find in the literature updates. It is possible to view the $\log(gf)$ vs $\Delta\log(gf)$ plots of individual elements, per cross-matched database, using the BRASS database: \url{brass.sdf.org}.

\begin{figure*}[!t]
\centering
\includegraphics[width=18.5cm]{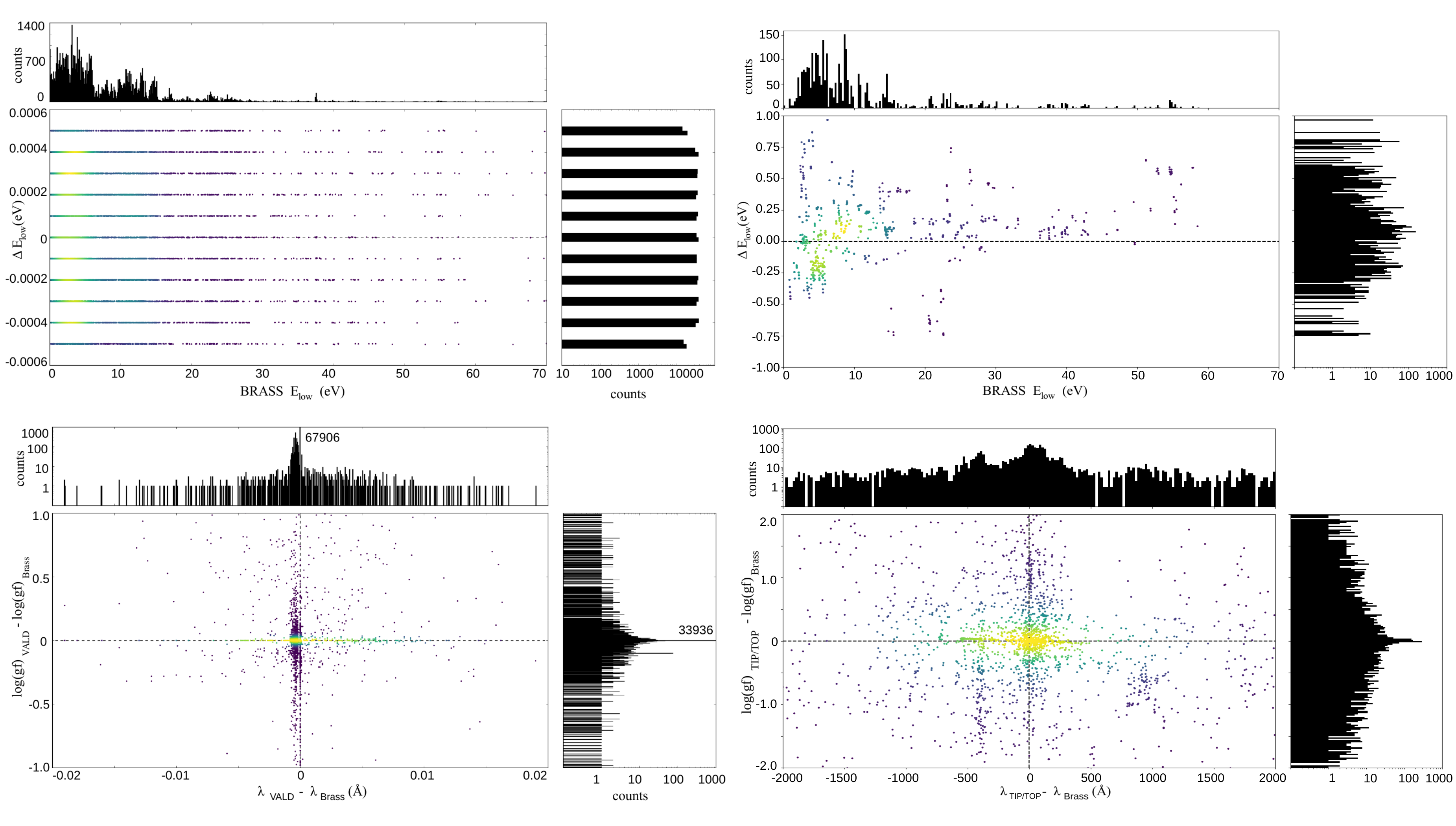}
\caption{ (Top left) $E_{i}$ vs $\Delta E_{i}$  for VALD3 vs BRASS.  (Top right) $E_{i}$ vs $\Delta E_{i}$  for TIPTOP vs BRASS. (Bottom left) $\Delta\lambda$ vs $\Delta \log(gf)$ for VALD3 vs BRASS. (Bottom right) $\Delta\lambda$ vs $\Delta \log(gf)$ for TIPTOP vs BRASS. }
\label{part67t}
\end{figure*}

\begin{figure*}[!t]
\centering
\includegraphics[width=18.5cm]{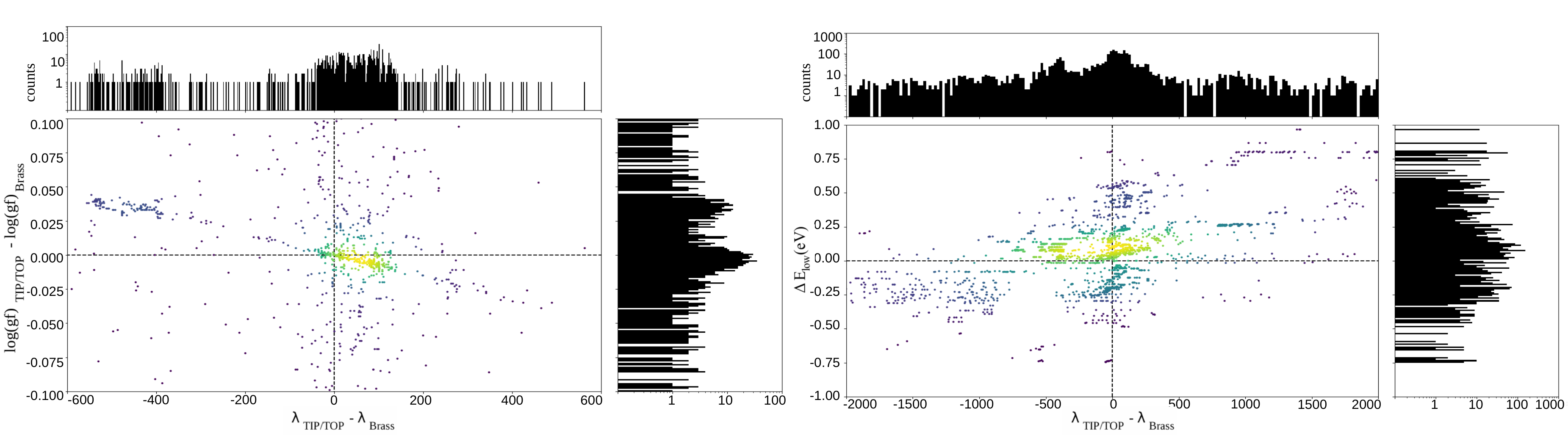}
\caption{ (Left) Smaller-scale correlations between $\Delta\lambda$ vs $\Delta \log(gf)$ for TIPTOP vs BRASS. (Right) $\Delta\lambda$ vs $\Delta E_{i}$ for TIPTOP vs BRASS. }
\label{part8t}
\end{figure*}

The top left panel of Figure~\ref{part5t} shows the $\lambda$ vs $\Delta\lambda$ for the BRASS and VALD3 non-parametric cross-match. We found that the differences in wavelength were reasonably well constrained. The majority of non-zero $\Delta\lambda$'s are well within $\pm$0.01\AA. Such small wavelength shifts will be unnoticeable in our benchmark spectra, which were obtained using the Mercator-HERMES and VLT-ESO-UVES \citep{uves} spectrographs, with maximum resolutions of R$\sim$85~000 and R$\sim$110~000 respectively. Within the $\pm$0.01\AA~range we observe two non-zero linear correlations between $\lambda$ and $\Delta\lambda$ for a small subset of lines. These correlations are caused by slight differences in precision of conversions, of vacuum-to-air wavelength or of units, within the source literature relative to the conversions used for the vast majority of VALD transitions. These correlations are visible in the top right panel of Figure~\ref{part5t} and cause the slightly negative skew of the histograms in the bottom left panel of Figure~\ref{part67t}. The $\lambda$~vs~$\Delta\lambda$ plots are distributed similarly for NIST and so the $\lambda$~vs~$\Delta\lambda$ plot of NIST is omitted. The $\lambda$~vs~$\Delta\lambda$ distribution for SpectroWeb is shown in the bottom left panel of Figure~\ref{part5t} and exhibits a reasonably constrained wavelength distribution within the 0.1~\AA~range used for the parametric cross-match. The bottom right panel of Figure~\ref{part5t} shows the $\lambda$~vs~$\Delta\lambda$ distribution for TIPTOP. We find that the $\lambda$ values of TIPTOP can disagree with more recent $\lambda$ values by as much as 2000~\AA. The diagonal lines are a result of comparing all cross-matched fine-structure BRASS transition wavelengths against the single mean TIPTOP multiplet wavelength. The large wavelength differences occur for singlet, complete multiplet, and incomplete multiplet cross-matches. As the goal of TIPTOP was to produce opacities, where wavelength accuracy is much less important than the line strengths and line list completeness, this is an unsurprising result.

Figure~\ref{part67t} (top left panel) shows the $E_{i}$ vs $\Delta E_{i}$ for the BRASS and VALD3 cross-match. The  $\Delta E_{i}$ values, and similarly $\Delta E_{k}$ values, are no larger than $\pm$0.0005~eV with a precision level of 0.0001~eV. The NIST cross-match and SpectroWeb cross-match both show the same phenomena as the VALD cross-match with no other additional trends, and so their plots have been omitted. The majority of transitions also have a lower energy level of $E_{i}$ between 0~-~20~eV, otherwise we do not find any obvious systematic differences or effects other than the precision of the energy levels. The top right panel of Figure~\ref{part67t} shows the $E_{i}$ vs $\Delta E_{i}$ distribution for TIPTOP. We find that the energy levels can vary by as much as 1~eV. These differences are due to significant energy level revisions and calculation improvements over the past decade. We find a slightly positive skew in the histogram of $\Delta E_{i}$, however the distribution of $E_{i}$ values is consistent with the other database cross-matches.

Figure~\ref{part67t} (bottom left panel) shows $\Delta \log(gf)$ vs $\Delta\lambda$ for the BRASS and VALD3 cross-match. We do not find any obvious systematics between $\Delta \log(gf)$ and $\Delta\lambda$ other than the expected $\Delta \log(gf)$ and $\Delta\lambda$ distributions shown in Figures~\ref{part4} and \ref{part5t}. This also applies to the SpectroWeb cross-match and NIST cross-match and so these plots have been omitted as they do not provide further information on correlations. The bottom right panel of Figure~\ref{part67t} shows the $\Delta \log(gf)$ vs $\Delta\lambda$ for the BRASS and TIPTOP cross-match. We find a similar distribution for the BRASS and TIPTOP cross-match, however we find three additional systematic trends in the TIPTOP cross-match: 1) A roughly vertical offset of points at around $-$~450~\AA~caused by \ion{C}{I}. 2) A cluster of points located around 1000~\AA~caused by the two lower energy levels, 4p~$^3$P$^{\circ}$ and 4p~$^3$D$^{\circ}$, in \ion{Ca}{I}. 3) A weak linear dependence between $\Delta\lambda$ and $\Delta\log{gf}$ for \ion{C}{III} and \ion{O}{III}, shown in the left panel of Figure~\ref{part8t}. 

Figure~\ref{part8t} (right panel) shows $\Delta\lambda$ vs $\Delta E_{i}$ for the TIPTOP cross-match. We find an overall positive correlation between $\Delta\lambda$ and $\Delta E_{i}$, in addition to many horizontal structures caused by the distribution of lower energy levels amongst the species of TIPTOP. The thick horizontal line, centred at roughly 1000~\AA, is caused by the two \ion{Ca}{I} 4p~$^3$P$^{\circ}$ and 4p~$^3$D$^{\circ}$ lower energy levels. The $\Delta\lambda$ vs $\Delta E_{i}$ figures for other databases do not reveal any correlations, due to the extremely constrained $E_{i}$ values, and so have been omitted.

\subsection{Differences between VAMDC-VALD and VALD3}

On 2016-03-14 we retrieved VALD3 via the VAMDC portal. For clarity in this section we shall refer to this data retrieval as VAMDC-VALD and the original VALD3 retrieval as VALD3-native. The VAMDC-VALD retrieval followed the same retrieval configurations as the other retrievals in Section 3: atomic transitions in the wavelength range 4200~-~6800~\AA~with ions of up to 5+. VAMDC-VALD was also cleaned of duplicate transitions and cross-matched against BRASS in the same manner as both VALD3-native and NIST. We found that the VAMDC-VALD cross-match did not produce the same results as the VALD-native cross-match.

Discussion with the VALD team {\color{blue}(Ryabchikova T., priv. comm. 2017)} revealed that VAMDC-VALD and VALD3-native should be identical, and that there had been a conversion issue with the electronic configurations when retrieving through VAMDC, leading to a number of transitions being assigned incorrect electronic configurations. The incorrect electronic configurations led our non-parametric cross-match to match entries that belonged to different physical transitions and thus we observed an increase in $\Delta \log(gf)$ scatter when producing our $\log(gf)$ vs $\Delta \log(gf)$ plots for VAMDC-VALD. The VALD team have confirmed that this error has now been corrected. This example provides a clear case for the importance of clarity when dealing with atomic data, be it detailed version history at the database and provider level, or detailed reporting of data provenance and origin at the user level.

\section{Duplicate transitions in repositories}

In addition to cross-matching different line lists and repositories against each other, the non-parametric cross-match can be used to find multiple occurrences of the same physical transition within a single repository, which we refer to as duplicated transitions. Non-parametric cross-matches were performed for the following line lists and repositories:

\begin{enumerate}[(1)]
\item BRASS against BRASS 
\item NIST against NIST
\item VALD against VALD 
\end{enumerate}

Table~\ref{duplicates} shows the number of initial potential duplicated transitions we found using our non-parametric cross-match. We found on the order of 10$^3$ duplicates for VALD, in the wavelength range of 4200-6800\AA~and for ions up to 5+, while for NIST we found 388 duplicate transitions. All of the potential duplicated NIST transitions contained either an empty upper or lower electronic configuration, excluding parity information, resulting in an identical configuration string of either odd or even parity, we therefore cannot conclusively say these transitions are real duplicates and thus they are retained. This, however, is not the case for the VALD lines that do contain full upper and lower configurations. The non-parametric cross-match did not initially distinguish between hyperfine transitions, between isotopic transitions and between different types of forbidden transitions. To correctly identify and flag duplicate transitions in VALD we must account for these three issues. Unfortunately, neither VALD nor NIST contain information on hyperfine structure, so we must manually account for this in our cross-match. Isotopic and transition type information are present in the databases. In the case of VALD both isotopic and transition type information are reported within the reference information.

Hyperfine structure refers to small shifts and splittings in the energy levels of atoms, molecules and ions, due to interaction between the state of the nucleus and the state of the electron clouds. Elements with an even number of both protons and neutrons have a nuclear spin of zero and do not exhibit hyperfine structure. Using this principle we can easily determine a number of transitions that cannot be of hyperfine origin and thus remain flagged as potential duplicates. For the remaining transitions, for which we could not rule out hyperfine structure using zero nuclear spin, we consulted the corresponding references to try and determine if the transition is of hyperfine origin. We checked each reference, corresponding to a potentially duplicated transition, for any mention of hyperfine structure. If a given reference accounted for hyperfine structure in their calculations or observations then all transitions belonging to that reference were flagged as potential hyperfine lines and no longer considered as a duplicate entry.

VALD transitions include isotopic information in the final part of the reference string written as Z~or~(A)Z, where Z denotes the element and A denotes the isotopic mass number if applicable. Additionally VALD also employs a new transition type flag {\color{blue}(Stempels E., priv. comm. 2017)}, located in the reference information, where:

\begin{enumerate}[A -]
\item denotes an autoionising transition
\item denotes a forbidden E2 transition
\item denotes a forbidden M1 transition
\item denotes a transition between polarization levels (P)
\end{enumerate}

We found that a number of potentially duplicated transition pairs were actually forbidden transitions each belonging to an electric quadrupole, E2, and a magnetic dipole, M1, type transition. According to transition selection rules it is possible for both E2 and M1 forbidden transitions to occur between the same energy levels so these transitions are not actually duplicates unless $|J_k - J_i| = 2$ . After subtracting possible hyperfine structure, isotopic, and E2-M1 forbidden pair transitions we compile the final number of duplicated transitions shown in Table~\ref{duplicates}.

\begin{table}
\centering
\caption{Number of duplicates found in NIST and VALD-based repositories. Initial matches are based purely on identical upper and lower configurations and upper and lower \textit{J}-values for a given ion. The literature checked matches have been cleaned of potential hyperfine, isotopic, and forbidden E2-M1-pair transitions. }
\begin{tabular}{c c c}
            \hline
            \hline
            \noalign{\smallskip}
            Repository      &  Initial matches & Final matches \\
            \noalign{\smallskip}
            \hline
            \noalign{\smallskip}
            BRASS		& 3589 & 2982 \\            
            VALD3		& 6351 & 3394 \\
            NIST 	& 388 & 0~$^a$  \\            
            \hline            
\end{tabular}
\tablefoot{$^a$All NIST duplicates are lines with either an undesignated upper or lower energy level and so we cannot conclude they are duplicate transitions. }
\label{duplicates}
\end{table}

\begin{table}
\centering
\caption[]{The ionic distribution of duplicated transitions present in our VALD3 (2016-05-26) retrieval. The majority of duplicated transitions are \ion{S}{I} lines \citep{klines} present in the source literature. }
\begin{tabular}{l r}
            \hline
            \hline
            \noalign{\smallskip}
            Ion in VALD3      &  Number of duplicates\\
            \noalign{\smallskip}
            \hline
            \noalign{\smallskip}
            
            \ion{Al}{II}		& 35\\            
            \ion{C}{IV}		& 2\\ 
            \ion{Cl}{II}		& 6\\ 
            \ion{Cu}{I}		& 10\\
            \ion{F}{II}		& 35\\
            \ion{Fe}{I}		& 37\\
            \ion{Fe}{II}		& 40\\
            \ion{Mo}{I}		& 1\\
            \ion{Rb}{I}		& 1\\
            \ion{S}{I}		& 3181\\
            \ion{Si}{III}		& 16\\
            \ion{Ti}{I}		& 30\\
            \hline    
            \noalign{\smallskip} 
            total duplicate lines & 3394 \\
            \hline     
\end{tabular}
\label{duplicatesvald}
\end{table}
Table~\ref{duplicatesvald} shows the ionic distribution of the duplicated lines for VALD. The majority of duplicate lines are \ion{S}{I} lines from \citet{klines}. We find that all VALD3 duplicate transitions originate from their respective sources rather than issues with the VALD compilation. The duplicates have been removed of hyperfine transitions and isotopic transitions. The full duplicate tables, including a separate table for the \ion{S}{I} lines, are available in both pdf and machine-readable form at \url{brass.sdf.org}.

Within the initial BRASS atomic line list we found duplicates originating from within the source references and a few duplicates caused by compilation issues when newer transitions are merged into the line list. While we may be able to find duplicate transitions in BRASS, it is beyond the scope of this paper to determine which transition is the correct one and should be retained. For the purposes of BRASS we have chosen to retain the weakest line, with the smallest $\log(gf)$, in order to minimise the impact of these dubious transitions on synthetic spectrum calculations.

\section{Spectral synthesis of atomic data}

\subsection{Impact of uncertain $\log(gf)$ values on synthetic spectra}

\begin{figure*}
\centering
\includegraphics[width=16cm]{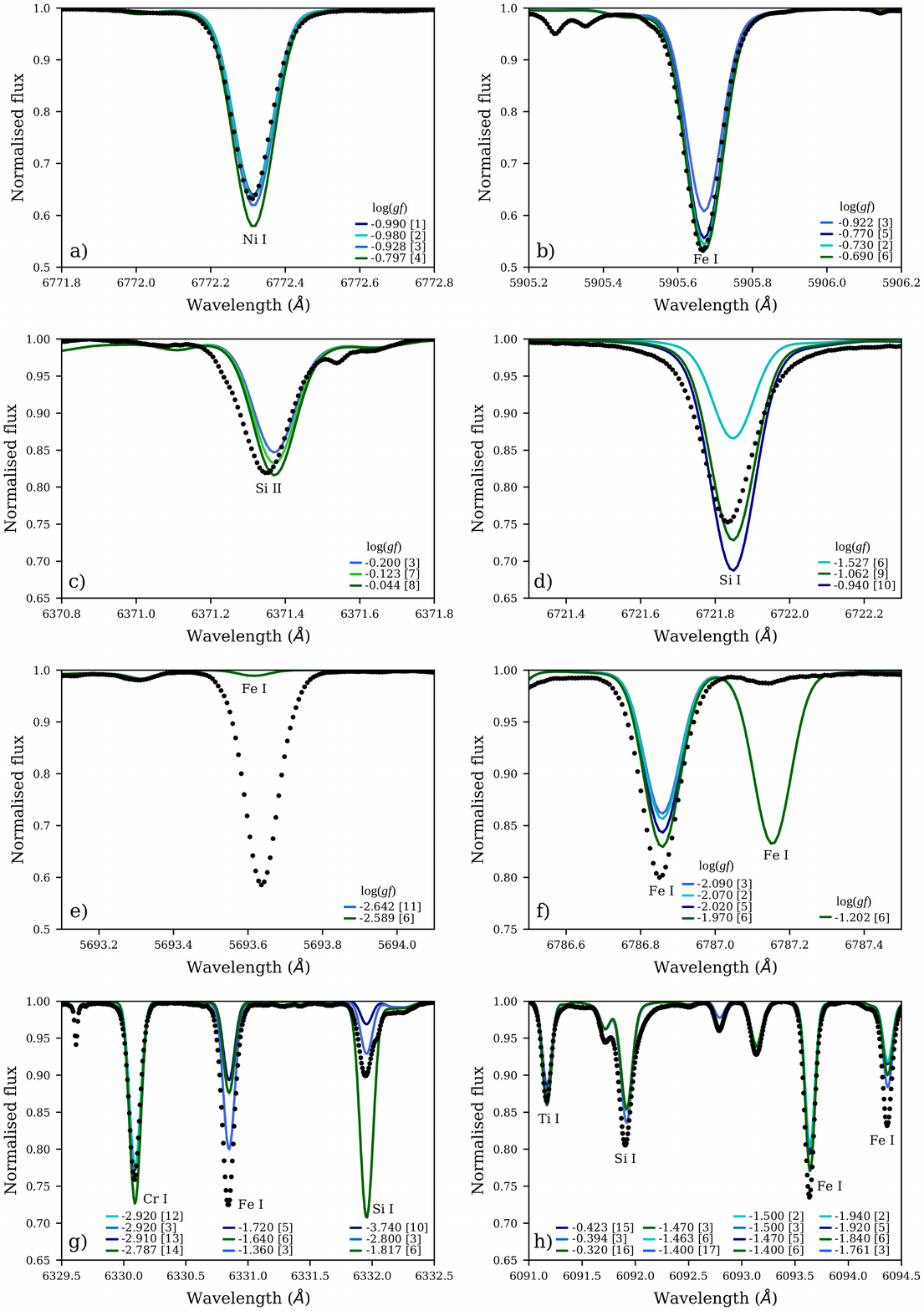}
\caption{ Examples of a number of different synthesised neutral and ionic lines, using our multiple cross-matched literature $\log(gf)$ and rest-wavelength values, compared against the observed Kitts Peak FTS solar flux spectrum (black). a) \& b) show examples where the majority of literature $\log(gf)$ values provide good fits to the observed line profile, with the exception of one value that systematically differs. c) shows a similar example but for \ion{Si}{II}, a species with relatively few usable clean lines in the visible spectrum. In addition, the central rest wavelength does not match the observed spectrum. d) shows an example where literature $\log(gf)$ value differences can reach as high as $\Delta\log(gf)~\approx~\pm$0.5~dex, and also a slight offset between observed and theoretical rest wavelengths. e) \& f) show examples of the so-called missing and unobserved lines. g) \& h) show a number of examples of lines with inaccurate $\log(gf)$ values when compared with stellar spectra. References can be found in Figure~\ref{loggf_synthesis2}.
              }
\label{loggf_synthesis}
\end{figure*}

Having completed the literature atomic transition cross-match we are now able to synthesise and inspect the impact of the literature scatter in $\log(gf)$ values that we observe in Figures~\ref{part1}, \ref{part2}, and \ref{part4}. Figure~\ref{loggf_synthesis} shows a number of examples where the literature $\log(gf)$ values show non-negligible differences. Adopting a certain $\log(gf)$ value can lead to systematic differences in $\log(gf)$, and thus line depth and abundances, which would remain hidden unless the atomic data are reported properly and compared against other literature work and values. Figure~\ref{loggf_synthesis} compares the solar Kitt Peak FTS flux spectrum of the Sun, described by \citet{solarfts}, against synthetic spectral calculations using cross-matched atomic data. The spectra have been synthesised using the TurboSpectrum spectral synthesis code \citep{turbospectrum}, \citep{plez2012}, 1D ATLAS9 atmospheric models \citep{atlas9}, and in conjunction with the opacity distribution functions of \citet{castelliopacity}. The solar abundances of \citet{solar2009} are adopted, and a microturbulent velocity of 1.1~km/s is used.

\begin{figure*}
\centering
\includegraphics[width=16cm]{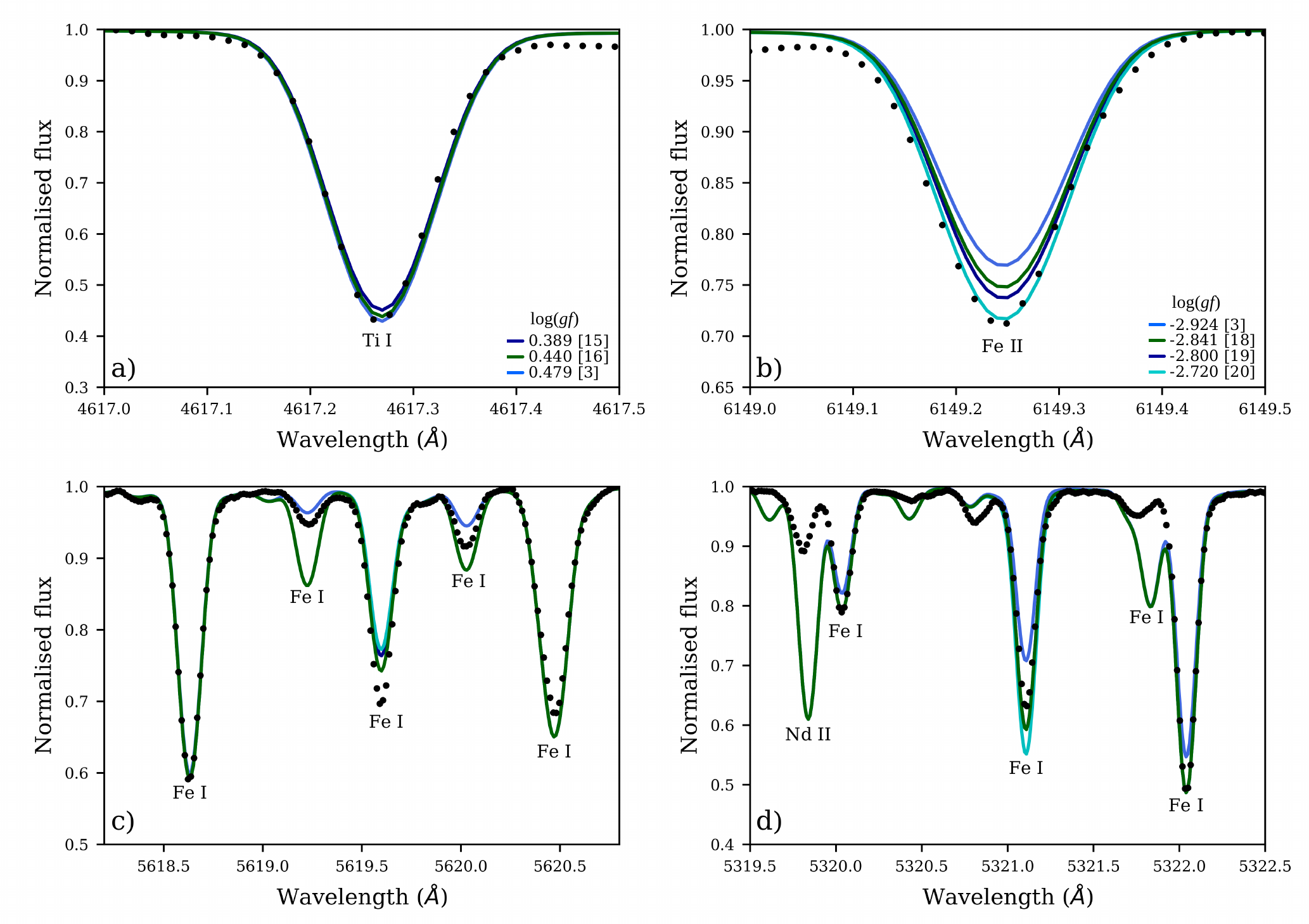}
\caption{ As in \ref{loggf_synthesis} but compared against the BRASS benchmark spectrum of 51~Peg (G2.5IV), observed using the Mercator-HERMES spectrograph (black). a) Multiple literature $\log(gf)$ values that all provide a good fit to the observed line profile. b) Only one literature $\log(gf)$ value out of four fits the observed line profile well. c) and d) show a number of \ion{Fe}{I} line $\log(gf)$ values with varying qualities of fitting to the observed profiles ranging from excellent fit to poor fit. $^{[1]}$\citet{mos1} $^{[2]}$\citet{FMW} $^{[3]}$\citet{spectroweb}  $^{[5]}$\citet{MRW}  $^{[7]}$\citet{chianti} $^{[8]}$\citet{mos8} $^{[9]}$\citet{mos9} $^{[10]}$\citet{mos10}  $^{[12]}$\citet{mos12} $^{[13]}$\citet{mos13} $^{[14]}$\citet{mos14} $^{[15]}$\citet{mos15} $^{[16]}$\citet{mos16} $^{[4,6,11,17]}$\citet{klines} $^{[18]}$\citet{Raassen} $^{[19]}$Same reference as [18] but reported by NIST to a limited precision. The $\log(gf)$ uncertainty is $\ge$50\% and calculated by NIST in the same manner as \citet{kramida2013} $^{[20]}$\citet{mos20}.
              }
\label{loggf_synthesis2}
\end{figure*}

\begin{figure*}
\centering
\includegraphics[width=16cm]{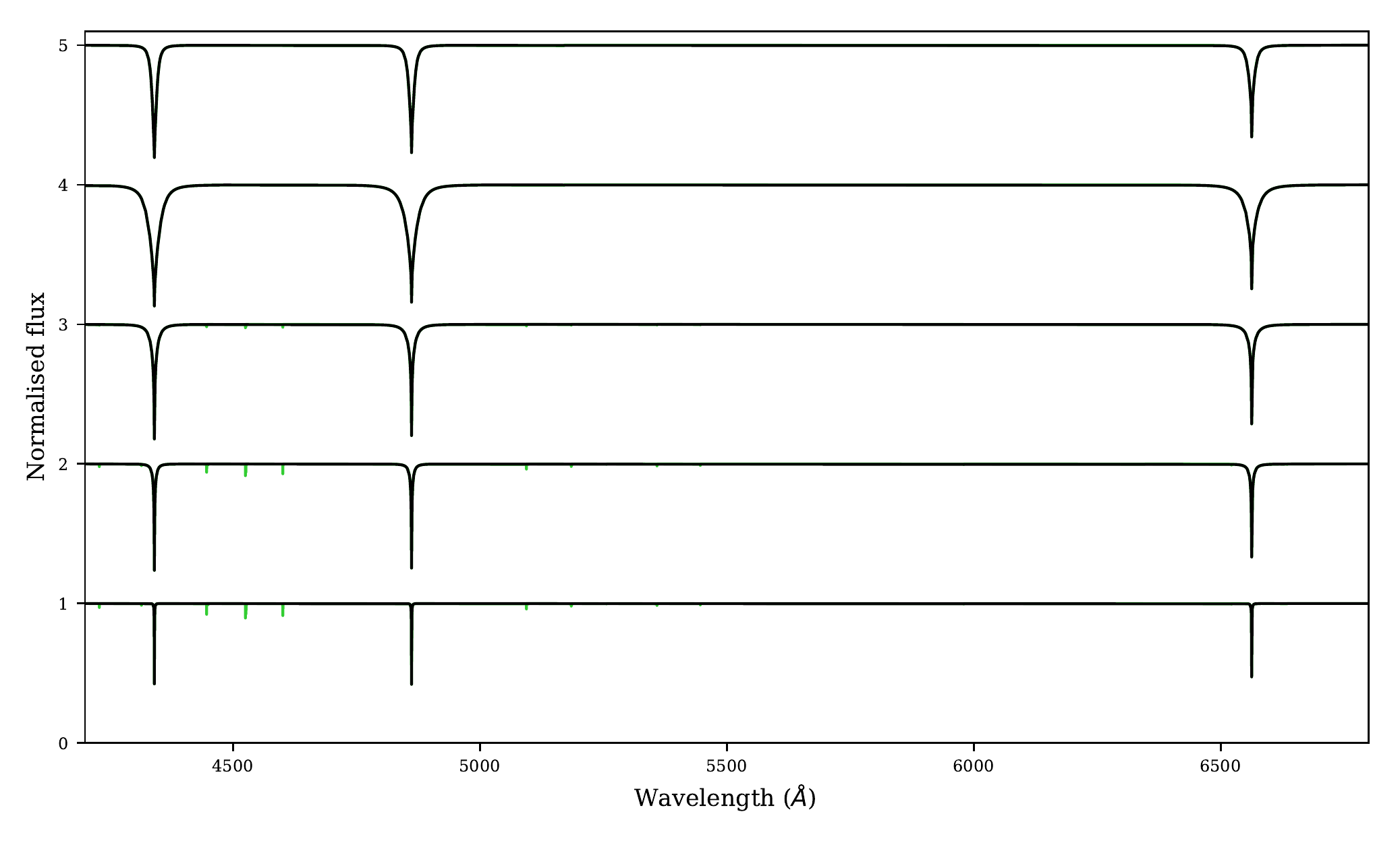}
\caption{ Synthetic spectra of only hydrogen lines (black) and duplicate transitions (green) synthesised for five BRASS benchmark stars: HR~7512 (B8~III), 68~Tau (A2~IV-Vs), Procyon (F5~IV-V), 51~Peg (G2.5~IV), and Arcturus (K1.2~III). Most duplicate lines produce a maximum depth, for the given spectral parameters, of less than 1\% of the continuum in this work. A small number of Fe duplicates produce maximum depths, at cooler stellar parameters, of at most 10\% of the continuum.
              }
\label{non_s1_duplicates}
\end{figure*}

Panels a) and b) of Figure~\ref{loggf_synthesis} show examples of a \ion{Ni}{I} line and a \ion{Fe}{I} line, at 6772.32~\AA~and 5905.67~\AA~respectively, for which the majority of $\log(gf)$ values are in agreement with each other and the solar spectrum. However each line also has a literature $\log(gf)$ value that disagrees with all other occurrences by as much as 0.2~dex. We find that for the \ion{Ni}{I} line this outlying value overestimates the observed line depth, whereas the \ion{Fe}{I} outlier underestimates the line depth. In these cases a careful review of the available atomic data would lead one to omit the outlying value. Unfortunately there are many examples where the majority of literature values disagree with each other making it hard to select an appropriate $\log(gf)$ value. In these cases one must either be familiar with the origin of the atomic data, or have access to quality assessments of atomic data, such as those that BRASS will provide.

Panels c) and d) show two such cases. The \ion{Si}{II} and \ion{Si}{I} lines, at 6371.37~\AA~and 6721.85~\AA~respectively, show examples where the literature $\log(gf)$ values disagree with each other to a larger extent than previously discussed. For the \ion{Si}{ii} line we find a maximum $\Delta\log(gf)$ of 0.15 dex, and for the \ion{Si}{i} line we find a much larger $\Delta\log(gf)$ of nearly 0.5 dex. Such differences in $\log(gf)$ values can have systematic consequences for abundance determinations, especially for species such as \ion{Si}{II} with few clean lines in the visible spectrum, as differences in adopted $\log(gf)$ values are inversely proportional to individual line abundance determinations for non-saturated spectral lines. For both these Si lines we also find slight offsets in wavelength between the theoretical and observed line profiles, stemming from rest wavelength values. BRASS shall produce quality of fit information, for all literature rest-wavelengths and $\log(gf)$ values of a vast number of deep unblended lines, in order to determine which literature values most accurately reproduce the BRASS benchmark spectra. The results of the assessment allow atomic data users to reliably build spectral line lists knowing the accuracy at which the selected data reproduce stellar observations.

The third set of spectra, Panels e) and f), show two common problems with stellar atomic line list compilations, the so-called missing lines and the unobserved lines. Missing lines are observed stellar lines or features that do not yet have known, measured, nor predicted atomic data. Unobserved lines are the inverse of this scenario, where the atomic data in question produce spurious features not observed in stellar spectra. The \ion{Fe}{I} line at 6787.155~\AA~is considered an unobserved line, and the feature at 5693.62~\AA~is considered a missing line. Unobserved lines originate from atomic calculations and thus their disagreement with observations could be due to either inaccurate wavelengths, predicted energy levels, or inaccurate $\log(gf)$ values. Typically, such lines are manually removed from line list compilations, but it is important that such cases are reported, so that they may be addressed in future atomic physics research. According to our current BRASS line list the missing feature at 5693.62~\AA~could actually be a \ion{Fe}{I} line with poorly fitting literature $\log(gf)$ values. To correctly fit this observed feature using the \ion{Fe}{I} line would require a significant increase in $\log(gf)$ value on the order of a few dex. To make such an identification would require a systematic assessment of the feature using multiple spectral types before one could correctly identify the feature as a transition belonging to a given atomic species. While such identifications are not currently the goals of the BRASS project, the data products available through BRASS, namely the extremely high-quality benchmark spectra, will certainly be a great boon for such research.

Panels g) and h) show a number of examples of varying levels of agreement in literature $\log(gf)$ values for spectral lines. The \ion{Cr}{I} and \ion{Ti}{I} lines, at 6330.091~\AA~and 6091.171~\AA~respectively, show further examples of reasonably well-fitted lines as long as the correct $\log(gf)$ value is adopted. The three \ion{Fe}{I} and \ion{Si}{I} lines, at 6330.848~\AA, 6093.643~\AA, 6094.373~\AA,~and 6091.919~\AA~respectively, show examples of $\log(gf)$ values that all systematically underestimate the line depth of the spectral lines. It is important to constrain such $\log(gf)$ differences against benchmark spectra, such as the Sun, so that more lines can be included in spectral synthesis calculations and future large-scale surveys. The \ion{Si}{I} line at 6331.956~\AA~is an important example where the literature $\log(gf)$ values disagree by as much as 2 dex. Such a large difference has a significant impact on spectrum modelling if atomic data are used blindly. The two \ion{Fe}{I} lines in Panel h) both belong to the same atomic multiplet, and as discussed in Section 3.1, the non-parametric cross-match can be used to build multiplet tables using the BRASS atomic line list. By combining the line quality assessment results with multiplet information, BRASS will investigate multiplets for potential systematic correlations and errors. 

The uncertainty on literature $\log(gf)$ values is a problem encountered in the spectroscopic analysis of all stars, not only the Sun. Figure~\ref{loggf_synthesis2} compares the retrieved and synthesised atomic data against a BRASS benchmark spectrum of 51~Peg~(G2.5IV). Details on the BRASS benchmark spectra and spectral parameters can be found in \citet{alexbrass}. Panel a) of Figure~\ref{loggf_synthesis2} shows a \ion{Ti}{I} line with multiple literature $\log(gf)$ values that are in good agreement with each other and the observed line profile. While there are a number of well fitting literature values, often used for diagnostic lines to determine atmospheric parameters, there are still a large number of observed lines with poor or uncertain literature data. Panel b) shows a \ion{Fe}{II} line with many literature occurrences but only one $\log(gf)$ value that reproduces the observed profile well. Ionic species tend to have poorer atomic data than neutral species due to the difficulty of producing such ionic species in the laboratory. Panels c) and d) show a number of \ion{Fe}{I} lines in 51~Peg. The \ion{Fe}{I} lines can vary significantly in quality of fit to the observed spectrum, with some $\log(gf)$ values fitting exceptionally well and others fitting rather poorly. The origin of such poor fits to the observed spectrum cannot be due to stellar parameter determination, and must instead be due to atomic data.

\subsection{Impact of duplicate transitions on synthetic spectra}

\begin{figure}
\centering
\includegraphics[width=9cm]{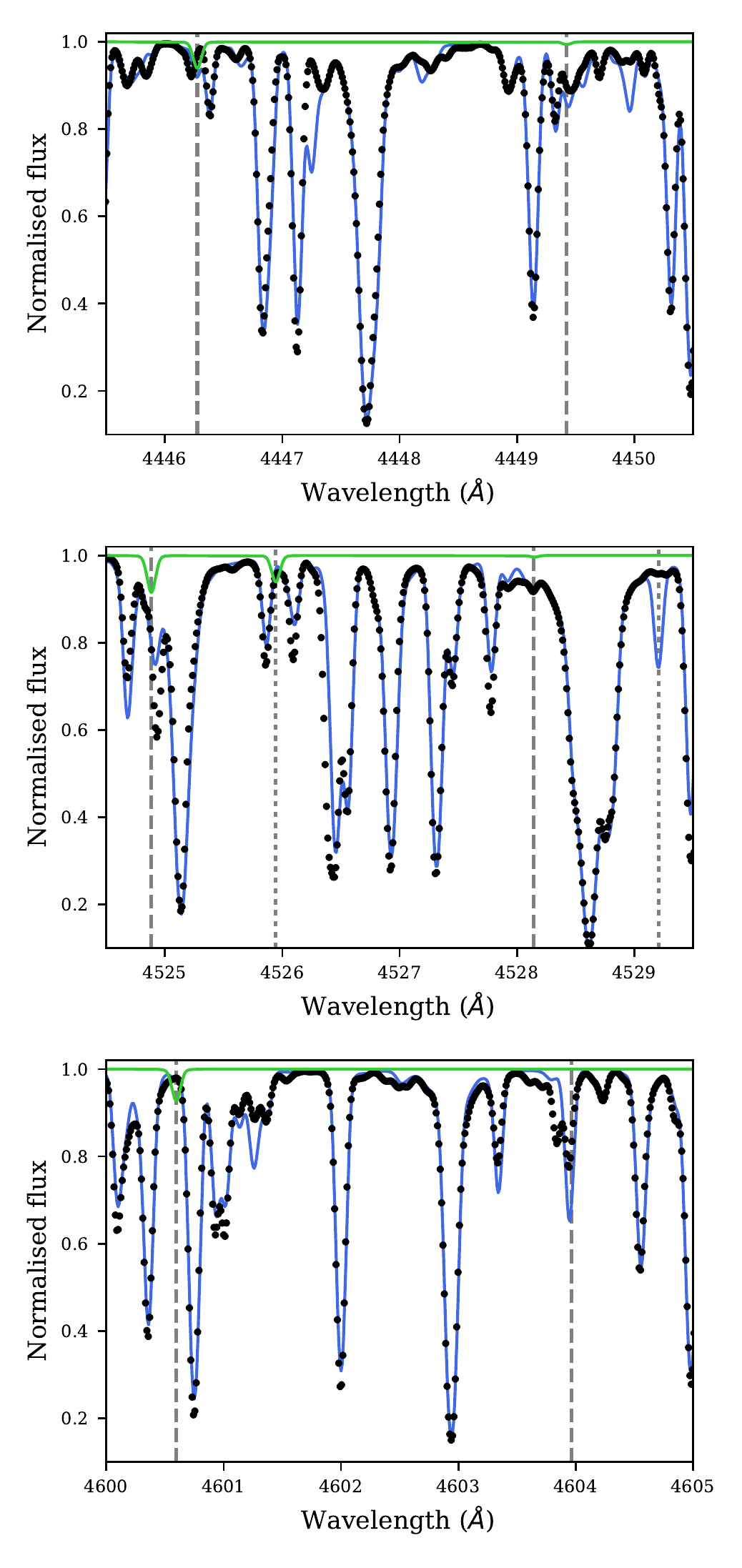}
\caption{ Comparison of the four strongest duplicate pairs present in the source literature. Duplicate lines profiles are shown in green, the duplicate-free BRASS spectrum is shown in blue, and the solar FTS flux spectrum, described by \citet{solarfts}, is shown in black. The four duplicate pairs rest wavelengths, shown by dashed lines, are as follows: (top) 4446.275\AA~\&~4449.424\AA, (middle, dashed) 4524.883\AA~\&~4528.144\AA, (middle, dotted) 4525.944\AA~\&~4529.206\AA, and (bottom) 4600.595\AA~\&~4603.966\AA.
              }
\label{non_s1_duplicates_comp}
\end{figure}

Figure~\ref{non_s1_duplicates} shows the synthesised duplicate lines, alongside the hydrogen lines, for five BRASS benchmark stars of B-, A-, F-, G-, and K- spectral type. Only a small number of duplicate lines are visible in our synthesis, typically Fe lines, and their impact on our stellar spectral synthesis is small when compared with other atomic and molecular contributions, as shown in Figure~\ref{non_s1_duplicates_comp} for the 4 strongest duplicate pairs. We find that for all benchmark stars the line depth of the strongest line in a duplicate pair does not exceed a value of $F/F_{Cont} = 0.9$. The strongest duplicate of each pair produces either an unobserved feature, suggesting that the atomic data of the line is wrong, or is heavily blended with other features making it difficult to assess the accuracy of the line.

Almost all of the duplicated \ion{S}{I} lines are autoionising transitions, according to the VALD transition type flags, which are not computed by the TurboSpectrum code. The few remaining \ion{S}{I} transitions are too weak to meaningfully impact our spectral synthesis calculations. Fortunately it appears that the duplicates present in the source literature are of negligible consequence for our spectral synthesis calculations, however they could still impact metal-rich stellar spectroscopy as well as atomic calculations. It is important that the duplicates are addressed.

\begin{figure}
\centering
\includegraphics[width=9cm]{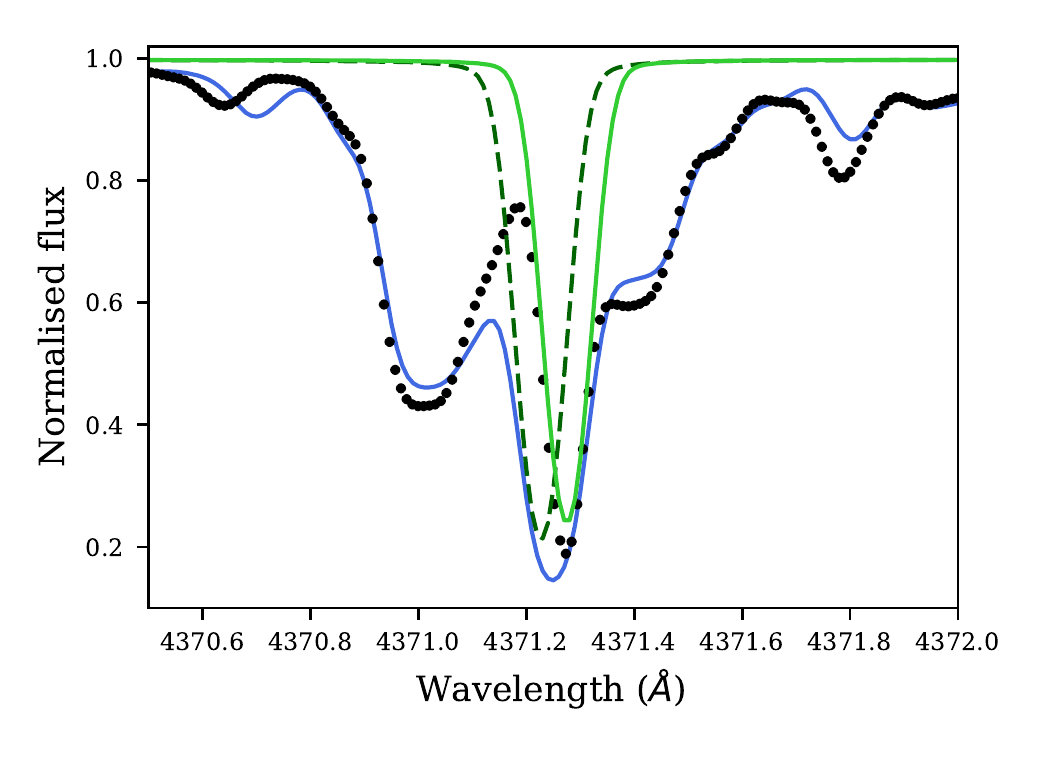}
\caption{ Two \ion{Cr}{I} duplicate transitions, introduced during the initial BRASS atomic line list compilation, synthesised for the Sun. The wavelength of the \ion{Cr}{I} line shown in dashed green is 4371.23~\AA, and the wavelength of the \ion{Cr}{I} line in solid green is 4371.28~\AA. Shown in black is the FTS solar spectrum described by \citet{solarfts} and shown in blue is a synthetic solar spectrum using the initial BRASS synthesis list including both \ion{Cr}{I} lines.
              }
\label{duplicatesynth}
\end{figure}

A small number of duplicates, present in the BRASS atomic line list, originate from the line list compilation of multiple sources, as opposed to being present in the source literature. Such compilation mistakes can have a significant impact on spectral synthesis and the non-parametric cross-match proves to be a useful tool for ensuring accuracy when compiling line lists. Figure~\ref{duplicatesynth} shows a \ion{Cr}{I} duplicate pair within our BRASS atomic line list and synthesised for the Sun. We find in this case that the inclusion of the \ion{Cr}{I} line at 4371.23~\AA~(dashed line), leads to an incorrect modelling of the solar FTS spectrum. Care must be taken to compile line lists as accurately and robustly as possible to prevent systematic errors from propagating through to future atomic quality assessment work. In this case the BRASS automatic line-blending investigations, discussed in \citet{mybrass}, would classify the \ion{Cr}{I} line as a heavily blended feature at $\lambda$~=~4371.25~\AA~and would be rejected from further quality assessment of available atomic data values. With the 4371.23~\AA~\ion{Cr}{I} line removed, the remaining \ion{Cr}{I} line can be used in quantitative spectroscopy.

\section{Summary and future work}

Input atomic data remains one of the main sources of uncertainty for spectral synthesis calculations, and it is vital that these uncertainties are constrained as quickly as possible. The Belgian repository of fundamental atomic data and stellar spectra will take the first steps towards removing systematic input atomic data errors from stellar spectroscopy. We shall critically evaluate the available literature data, associated with over a thousand prominent and unblended atomic lines, by synthesising each line and systematically comparing the calculated line profiles against several extremely high-quality observed spectra spanning the BAFGK spectral types. BRASS will provide all quality assessments, retrieved atomic data, and all extremely high-quality benchmark spectra, via a new interactive, graphical online database. In addition, BRASS will also provide $\sim$100 high-quality spectra, further sampling the BAFGK spectral parameter space, for use by the astronomy community. These spectra shall include $\sim$40 new hot radial velocity standard stars. The database shall take the novel approach of graphically offering atomic data and spectra together, providing atomic data parameters alongside synthesised atomic data and observed stellar spectra. 

As a first step in the BRASS project we have retrieved over 400~000 repository transition entries from VALD3, NIST ASD, SpectrW$^3$, TIPbase, TOPbase, CHIANTI and the SpectroWeb line lists. Atomic transitions were retrieved for neutral species and ions up to 5+ in the wavelength range of 4200-6800~\AA. 

All retrieved database transitions were homogenised and cross-matched against the BRASS atomic line list, composed of Kurucz and NIST v4.0 transitions, using two different methods: a parametric cross-match using constraints on wavelength and energies to match similar transitions for a given ion, and a non-parametric cross-match that uses electronic configurations to match transitions of the same physical origin for a given ion. All cross-matches accounted for fine structure, isotopic information, and transition type, but not hyperfine structure due to lack of hyperfine structure information. The inhomogeneities of electronic configuration nomenclature both between and within VALD and NIST are presented. Using our non-parametric cross-match we explore the differences between multiple occurrences of the same physical transitions across the literature. For literature distributions of $\lambda$~vs~$\Delta\lambda$'s, $E$~vs~$\Delta E$'s, $\Delta\lambda$ vs $\Delta E$'s, and $\Delta\lambda$ vs $\Delta \log(gf)$ values we find an absence of large-scale systematic correlations and the presence of small-scale conversion precision differences. We find significant scatter in cross-matched $\log(gf)$ values of up to 2~dex or more. This scatter has significant implications for spectroscopic analysis. All cross-matched atomic data, including the BRASS atomic line list, are available for download at \url{brass.sdf.org}. In addition, further $\log(gf)$ vs $\Delta\log(gf)$ plots for individual elements and databases are available online.

We used our non-parametric cross-match method to investigate the issue of duplicate transitions in the retrieved repositories. After accounting for hyperfine transitions, isotopic transitions, and E2-M1 forbidden transitions, we find a significant number of duplicated transitions, up to 2\% of our retrieved lines, in VALD3 and Kurucz lines. No duplicated transitions were conclusively found in our subset of NIST. We found that the duplicate transitions could be sourced back to the original work in 99\% of cases, meaning they were not produced by the repositories. The full tables can be found online at \url{brass.sdf.org}. The impact of compilation error duplicate transitions on synthetic spectral calculations is important for synthesis work, and thus we present a clear case for accurate line list compilations. 
We discuss the impact of scatter in literature $\log(gf)$ values on stellar spectral modelling. A number of examples are provided where correctly fitting $\log(gf)$ values must be carefully selected, assuming that there are any correctly fitting values at all, in order to reduce the risk of introducing systematic errors into spectral analysis. We provide examples showing disagreement in literature $\log(gf)$ of up to 2~dex, lines that are not present in observations, and lines that remain missing from theoretical calculations and line list compilations.

In the future we will systematically select strong unblended lines, using the cross-matched atomic data, for atomic quality assessments. We will critically compare and evaluate all literature occurrences of selected lines against extremely high-quality benchmark stellar spectra covering BAFGK spectral types. The cross-matched data, quality assessment results, synthetic spectra and observed spectra will be made publicly available in a new interactive online database at \url{brass.sdf.org}.

\begin{acknowledgements}
       We thank both Steven Shore and Alexander Kramida for their insightful comments and contributions towards the paper. We thank Peter van Hoof for the discussions and contributions towards the multiplet analysis of TIPbase and TOPbase. We also thank the atomic data producers and providers for their invaluable work towards improving the accuracy of stellar spectroscopy and the ease at which such vast quantities of data can be retrieved. The research for the present results has been subsidised by the Belgian Federal Science policy Office under contract No. BR/143/A2/BRASS. T.M is supported by a grant from the Fondation ULB. This work has made use of the VALD database, operated at Uppsala University, the Institute of Astronomy RAS in Moscow, and the University of Vienna. This work is based on observations made with the Mercator Telescope, operated on the island of La Palma by the Flemish Community, at the Spanish Observatorio del Roque de los Muchachos of the Instituto de Astrofísica de Canarias. This work is also based on observations obtained with the HERMES spectrograph, which is supported by the Research Foundation - Flanders (FWO), Belgium, the Research Council of KU Leuven, Belgium, the Fonds National de la Recherche Scientifique (F.R.S.-FNRS), Belgium, the Royal Observatory of Belgium, the Observatoire de Genève, Switzerland and the Thüringer Landessternwarte Tautenburg, Germany. 
\end{acknowledgements}

\end{document}